\documentclass[a4paper,11pt]{article}
\pdfoutput=1 %
\usepackage{jcappub} 

\usepackage[T1]{fontenc} 

\usepackage[utf8]{inputenc}

\usepackage{hyperref} 
\usepackage{url}
\usepackage{graphicx}
\usepackage{esvect}
\usepackage{graphics}
\usepackage{natbib}
\usepackage{mathrsfs}
\usepackage{blindtext}

\usepackage{comment}

\usepackage{lineno}

\usepackage[table]{xcolor}
\usepackage{multirow}

\usepackage[normalem]{ulem}
\usepackage{color}

\usepackage{soul}

\usepackage{supertabular}
\usepackage{tabularx}

\usepackage{academicons}

\def\units#1{~\hbox{$\,{\rm #1}$}}

\newcommand{\orcid}[1]{\href{https://orcid.org/#1}{\textcolor[HTML]{A6CE39}}}

\title{The FLUKA cross sections for cosmic-ray leptons and uncertainties on current positron predictions}

\author[a,1]{P.~De~La~Torre~Luque\,\orcid{0000-0002-4150-2539}\note{Corresponding authors.},}
\emailAdd{pedro.delatorreluque@fysik.su.se}
\author[b, c]{F.~Loparco\,\orcid{0000-0002-1173-5673}}
\author[b,1]{and M.~N.~Mazziotta\,\orcid{0000-0001-9325-4672}}
\emailAdd{mazziotta@ba.infn.it}

\affiliation[a]{The Oskar Klein Centre, Department of Physics, Stockholm University, AlbaNova\\
  SE-10691 Stockholm, Sweden}
\affiliation[b]{Istituto Nazionale di Fisica Nucleare, Sezione di Bari, via Orabona 4, I-70126 Bari, Italy}
\affiliation[c]{Dipartimento di Fisica ``M. Merlin" dell'Universit\`a e del Politecnico di Bari, via Amendola 173, I-70126 Bari, Italy}

\date{\today}

\abstract{
Cosmic-ray (CR) antiparticles have the potential to reveal signatures of unexpected astrophysical processes and even new physics beyond the Standard Model. Recent CR detectors have provided accurate measurements of the positron flux, revealing the so-called positron excess at high energies. However, the uncertainties related to the modelling of the local positron flux are still very high, significantly affecting our models of positron emission from pulsars and current dark matter searches. 

In this work, we report a new set of cross sections for positron and electron production derived from the {\tt FLUKA} code. We compare them with the most extended cross-section data-sets and show the impact of neglecting the positron production from heavy CRs.
Then, we review the most significant sources of uncertainties in our current estimations of the secondary positron flux at Earth and examine for the first time the impact of considering the spiral arm structure of the Galaxy in these estimations.
Finally, we provide state-of-the-art predictions of the local positron flux and discuss the limitations of our dark matter searches with positrons and difficulties to determine the contribution from pulsars to the positron flux at low energies.
}

\begin{document}
\maketitle
\flushbottom

\section{Introduction}
\label{sec:intro}
The potential of CR antiparticles (namely positrons, antiprotons and antinuclei) to reveal the presence of dark matter (DM) in the Universe has been discussed since the 80s~\cite{Silk_1984, Stecker_1985}. 
They offer a unique window to explore possible couplings with Standard Model (SM) forces~\cite{Leane_2018}. 
In particular, many of the best-motivated DM models predict the existence of Weakly Interacting Massive Particles (WIMPs) that naturally appear around the weak scale and can leave clear imprints in the spectra of Galactic CRs at GeV energies, where current experiments can provide sensitive measurements~\cite{DM_CRs}.

The measurement of the flux of Galactic CR antiparticles has already allowed us to place constraints on WIMPs that are in some cases even stronger than those obtained from accelerator searches or direct detection experiments for certain channels~\cite{PDG_review}.
However, there are important astrophysical uncertainties affecting these indirect searches: neither the transport of charged particles in the interstellar medium (ISM) nor the astrophysical mechanisms of injection of CRs by sources are well known. These uncertainties must be carefully investigated to correctly interpret the valuable information encrypted in the spectra of CRs measured at Earth. 

An interesting case that shows the impact of not considering carefully astrophysical uncertainties in our interpretation of the data was the exciting first measurement of the local positron spectrum by PAMELA~\cite{PAMELA:2008gwm}, which revealed an unexpected raise in the positron flux at high energies, the so-called \textit{positron excess}, pointing to a new source of positrons not considered before. This source of positrons was initially interpreted by many researchers as the signature of dark matter annihilating into leptons~\cite{Cirelli:2008jk, Ibarra_2009}. 
Nonetheless, the injection of electrons and positrons by pulsar wind nebulae (PWNs) was recognised to be a plausible explanation to the positron excess~\cite{SERPICO20122, Hooper_2009}, and currently there is compelling evidence showing that PWNs constitute the dominant source of positrons above 
$\sim 10$~GeV~\cite{Manconi:2020ipm}.
In turn, this contribution seems to represent at most a $\sim 15-20\%$ of the electron spectrum at $\sim300$~GeV.

In the standard scenario of Galactic CR propagation, electrons are basically accelerated and injected to the ISM by supernova remnants (SNRs), while positrons are generated by the interactions of protons and heavier CRs with the gas in the ISM~\cite{Gabici:2019jvz, Ginz&Syr, berezinskii1990astrophysics}. The positron excess requires also the injection of both electrons and positrons by PWNs.
Once they are released to the ISM, they interact with the Galactic magnetic field and plasma inhomogenieties and propagate through the magnetised Galactic halo~\cite{stephens1998cosmic} for long time~\cite{CarmeloBeB, moskalenko2000diffuse}, being subject to energy losses that may prevent them to travel for long distances. 

Currently, evaluations of the local positron flux are split into the study of PWNs emission, which dominates high energy region of the spectrum, and the study of the positron production from Galactc CR with the ISM gas, which dominate the low energy region.
The injection of positrons from pulsars is usually estimated integrating over the emission of an ensemble of nearby pulsars with averaged emission parameters (mainly  acceleration efficiency, spectral index of the power-law describing the emission, cut-off, etc.), frequently assuming a known spectrum of secondary positrons. The difficulty on measuring these parameters makes these estimations very uncertain, although the increasing number of high-energy gamma-ray measurements of PWNs are allowing us to have a better characterisation of their emission.

Similarly, the secondary positron production is largely uncertain, mainly due to our poor knowledge on the cross sections of production of these particles from CR interactions. This is a common problem in the studies of all secondary CRs (see, e.g. Refs.~\cite{Luque:2021joz, Luque:2021ddh, Korsmeier:2021brc, ICPPA_Pedro, GONDOLO2014175, GenoliniRanking, Tomassetti:2015nha}). CR leptons are produced from interactions of CR nuclei, mainly via the decay of unstable hadrons (essentially pions, kaons and neutrons below the GeV). Due to the amount of channels involved in their production, the broad energy range for which we need cross sections estimations and the lack of experimental data, there have been several attempts to develop a cross-section network purely from physical principles, relying on Monte Carlo simulations and event generators (e.g.~\cite{Bierlich:2022pfr, PSHENICHNOV2010604}) finding differences of up to a factor of $2$~\cite{Delahaye_PositronXSs} (see also Ref.~\cite{Koldobskiy}). 
Further uncertainties affecting our estimations on the flux of positrons (and electrons) at Earth are those on the Galactic magnetic field, interstellar photon fields, the gas density distribution and the Solar modulation, whose effect is usually naively approximated below a few GeV. 

This work aims at computing the full cross-section network of electrons and positrons using the {\tt FLUKA}~\cite{Ferrari:2005zk, FLUKA1, Arico:2019pcz} code and compare it with the most popular data-sets available.  Then, we implement these cross sections into the {\tt DRAGON2} code~\cite{DRAGON2-1, DRAGON2-2} and evaluate the main sources of uncertainties in the estimation of the local positron spectrum. This constitutes a follow-up of our recent work~\cite{delaTorreLuque:2022vhm}, where we computed the full network of nuclear spallation cross sections (including all isotopes up to Z=26) and analysed the predicted spectra of the secondary CRs B, Be and Li.
This paper is organized as follows: in section~\ref{sec:FlukaXS} we describe the calculations of electron and positron cross sections performed with the {\tt FLUKA} code and report them compared to the most up to date data-sets available and parameterisations. Then, in section~\ref{sec:Uncerts}, we explore the main sources of uncertainties affecting our predictions of secondary positrons at Earth, remarking the uncertainties related to the gas distribution, as well as the energy losses due to the turbulent component of the magnetic field. Finally, in section~\ref{sec:Discussion} we discuss the impact that these uncertainties have for both the determination of the injection spectrum of lepton pairs by PWN and our current searches for leptophilic dark matter.

\section{The Fluka cross sections for cosmic-ray leptons}
\label{sec:FlukaXS}

{\tt FLUKA}\footnote[1]{\url{http://www.fluka.org}.} is a general purpose tool that can be used to calculate the interactions of particles and their transport in arbitrarily complex geometries, including magnetic fields. The detailed treatment of hadronic interactions in a broad energy range makes it very suitable for its application in astroparticle physics studies (see, e.g.~\cite{Battistoni:2008hga, Andersenarticle, Heinbockel:2011zz, Tusnski:2019rpd, FlukaSun, Mazziot, delaTorreLuque:2022vhm}), raising high expectations in the CR community. A complete review of the recent developments implemented in {\tt FLUKA} and its current status will be published soon.

Remarkably, the high-energy FLUKA models have been deeply revised in the last years in order to improve predictions, and to better match experimental data on the production of intermediate hadronic resonances. Particularly, hadronic cross section fluctuations have been recently implemented in the code, improving emitted particle multiplicity distributions, and allowing for the first time to compute quasi-elastic and absorption cross sections directly from the built-in FLUKA Glauber model. A thorough revision and extension of the latter now allows the computation of reliable hadron-nucleus cross sections up to energies of $10^{20}$~eV. Detailed information on these models can be found in our previous work~\cite{delaTorreLuque:2022vhm} and its related publications and references can be found in the FLUKA webpage~\footnote{\url{http://www.fluka.org/fluka.php?id=publications&mm2=3}.}.
At the same time, several developments in the sub-GeV - few GeV energy range have improved the agreement of simulation results with experimental data.
A new model for coherent and incoherent (quasi-elastic) hadron-nucleus interactions has been developed and it is now applied to protons and neutrons up to $200$~MeV, with parameters optimized against available experimental data and evaluations taken from the JENDL-4 and ENDF/B-VIIIR0 nuclear data-bases~\cite{JENDL, Brown20181short}. 
The same approach is used for all hadrons above $1$~GeV, accounting also for quasi-elastic interactions at high energies, which are now explicitly described and no longer treated as a sub-case of coherent elastic interactions.  
The remarkable agreement between data and the {\tt FLUKA} predicted yield of pions, the main producer of CR leptons, can be seen in Fig.~2 of Ref.~\cite{Mazziot}. More detailed comparisons of the interactions and production of pions, kaons and other resonances important for the production of CR leptons can be found in Ref.~\cite{Cerutti_XSCR}.

In our companion work~\cite{delaTorreLuque:2022vhm}, we computed inelastic and inclusive cross sections of all stable isotopes from $^1$H to $^{60}$Fe impinging on helium and hydrogen as targets from $1 \units{MeV/n}$ to $35 \units{TeV/n}$, using $176$ bins equally spaced in logarithmic scale. We use these sets of simulations to obtain the full cross-section network of production of electrons and positrons. In addition to the prompt production of CR leptons, all the resonances and unstable nuclei incorporated in the {\tt FLUKA} code (see \url{http://www.fluka.org/content/manuals/online/5.1.html}) that produce leptons, either from direct decay or from a multi-step decay (e.g. the $\eta$ particle, that decays into pions $\sim70\%$ of the times, or the $\Delta$ resonance), contribute to their production and are considered. 
The final product of this chain of computations is a set of cross section tables for positron and electron production that can be directly used by any public CR propagation code. 

\begin{figure*}[!t]
    \centering
    \includegraphics[width=0.49\textwidth]{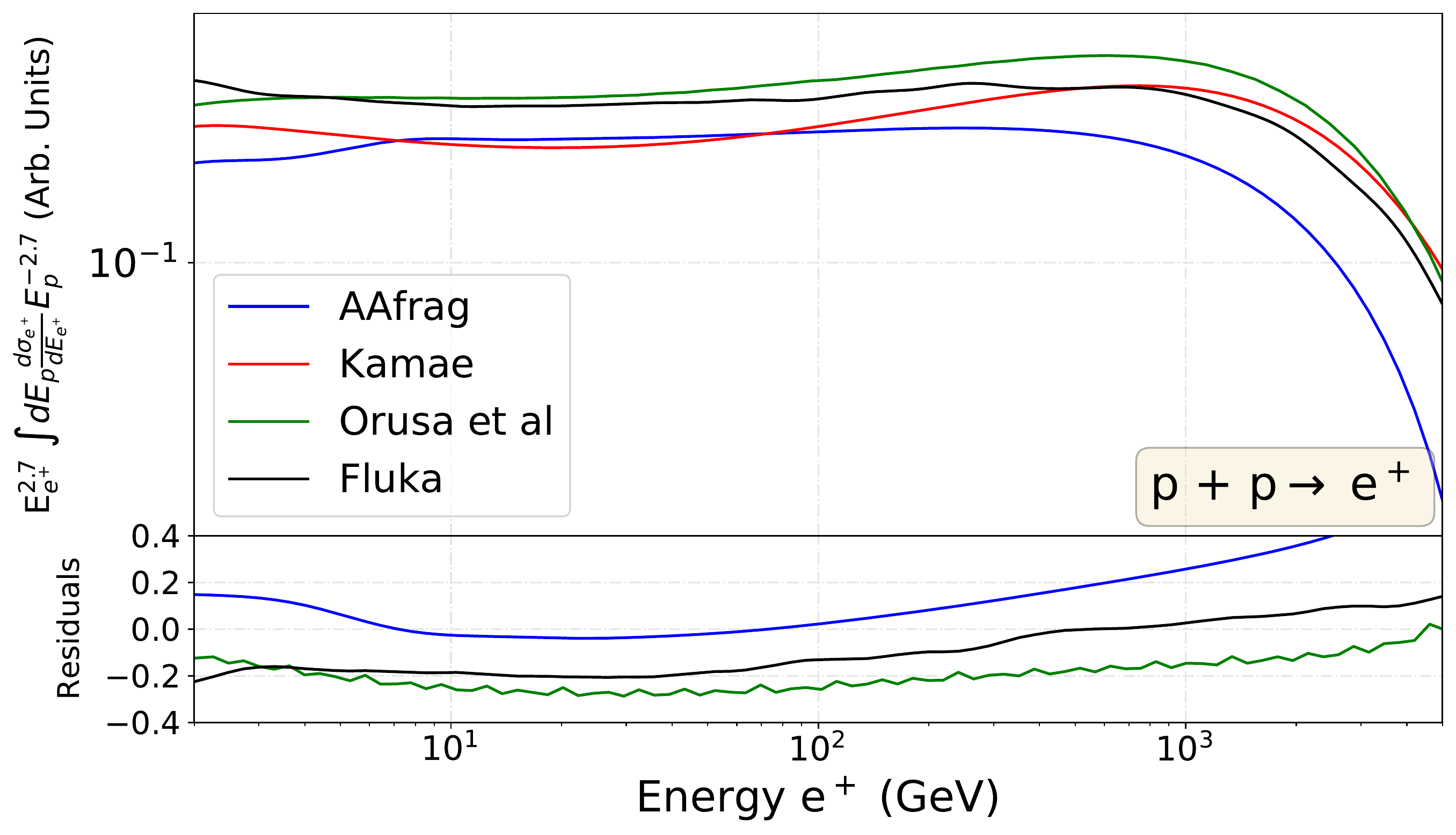}
    \includegraphics[width=0.49\textwidth]{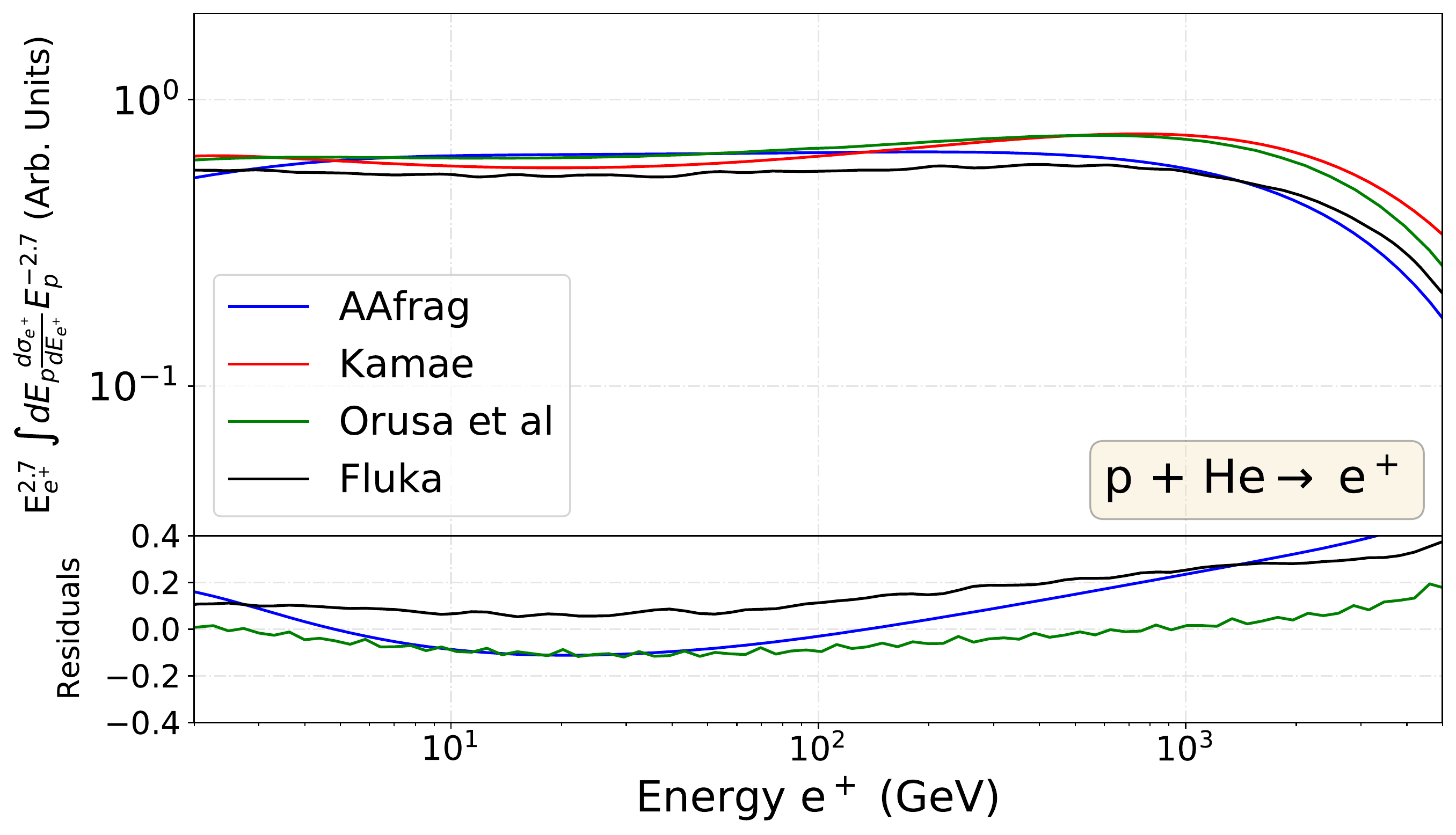}
    \includegraphics[width=0.49\textwidth]{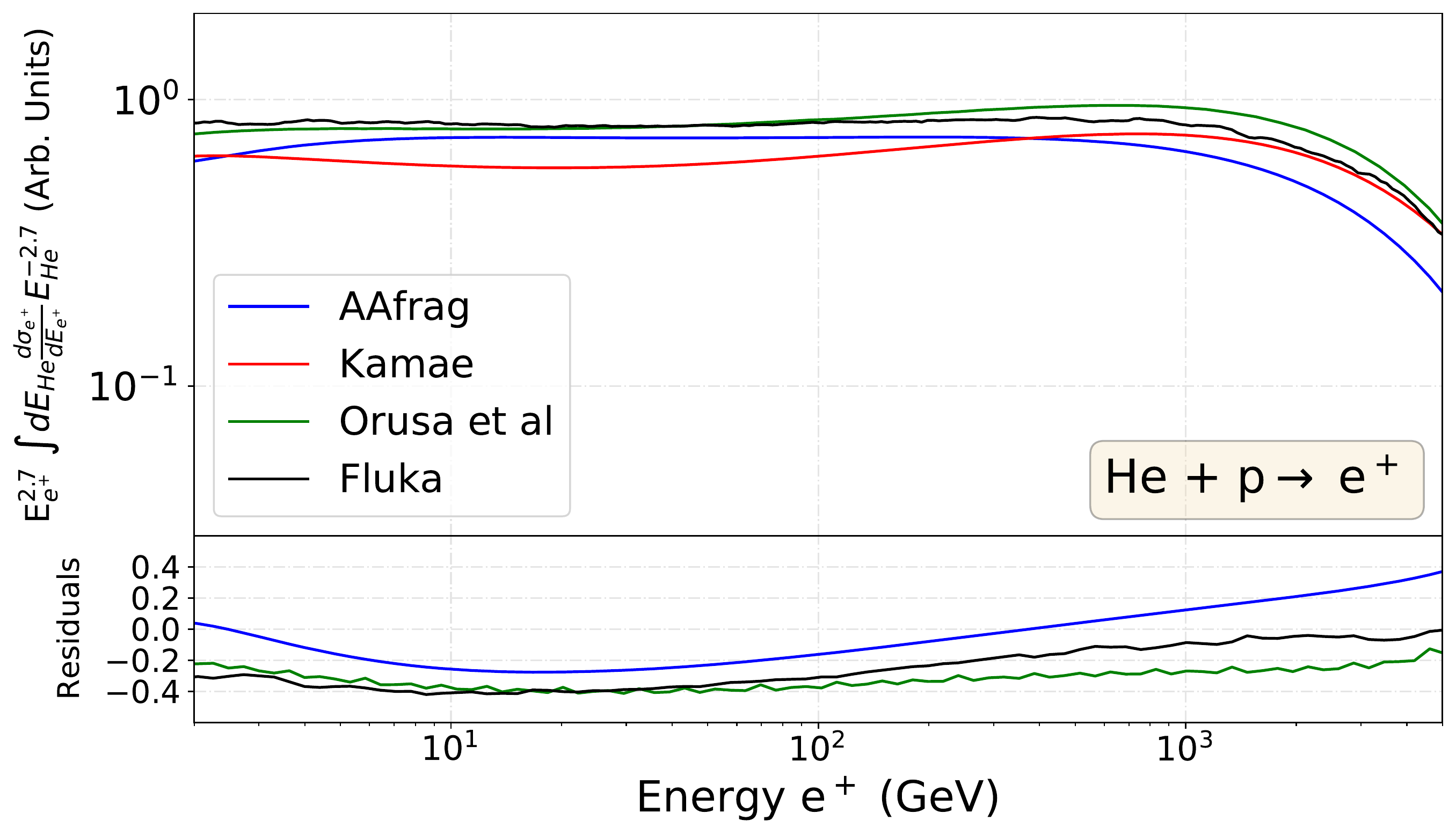}
    \includegraphics[width=0.49\textwidth]{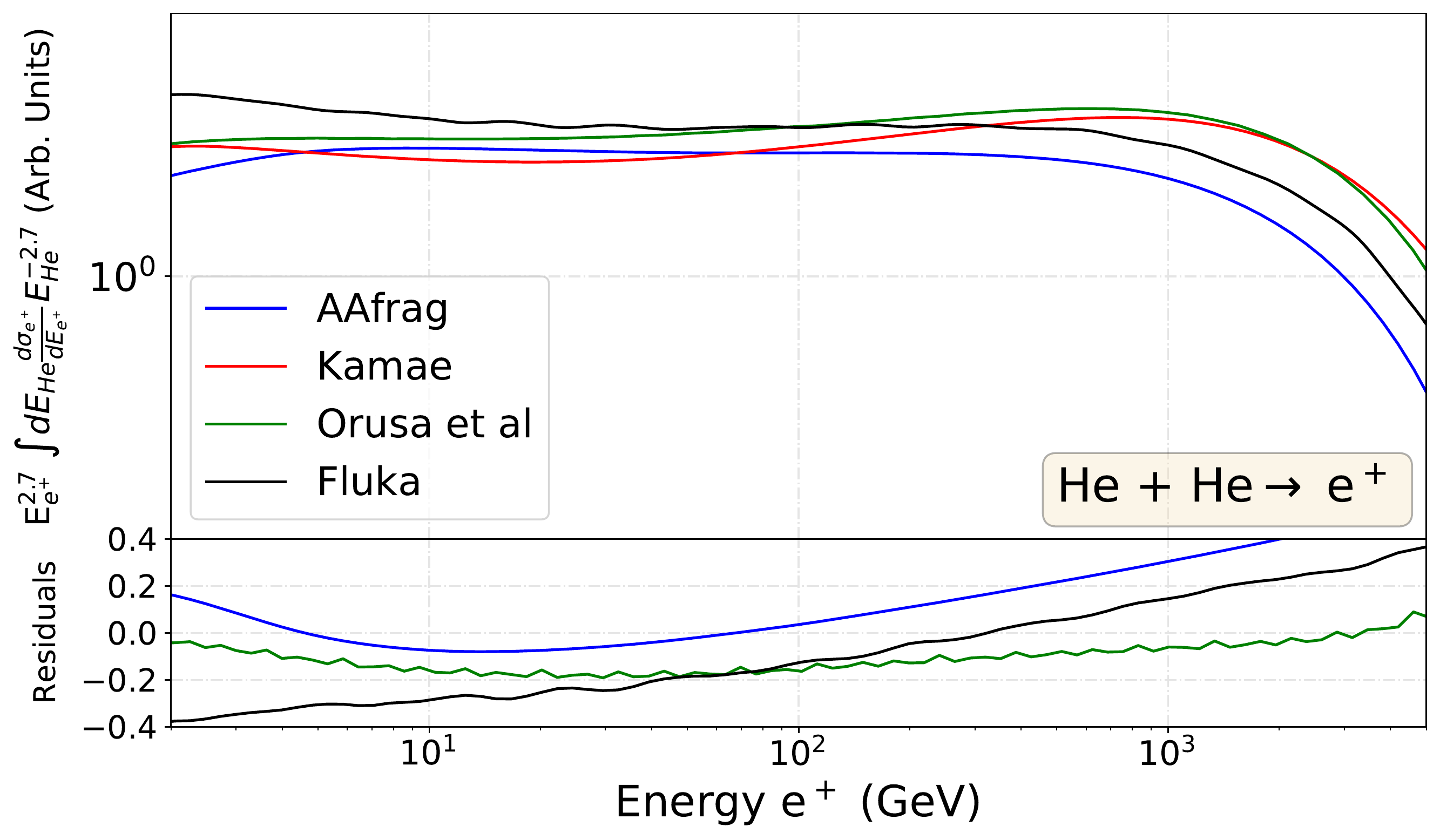}
    \includegraphics[width=0.49\textwidth]{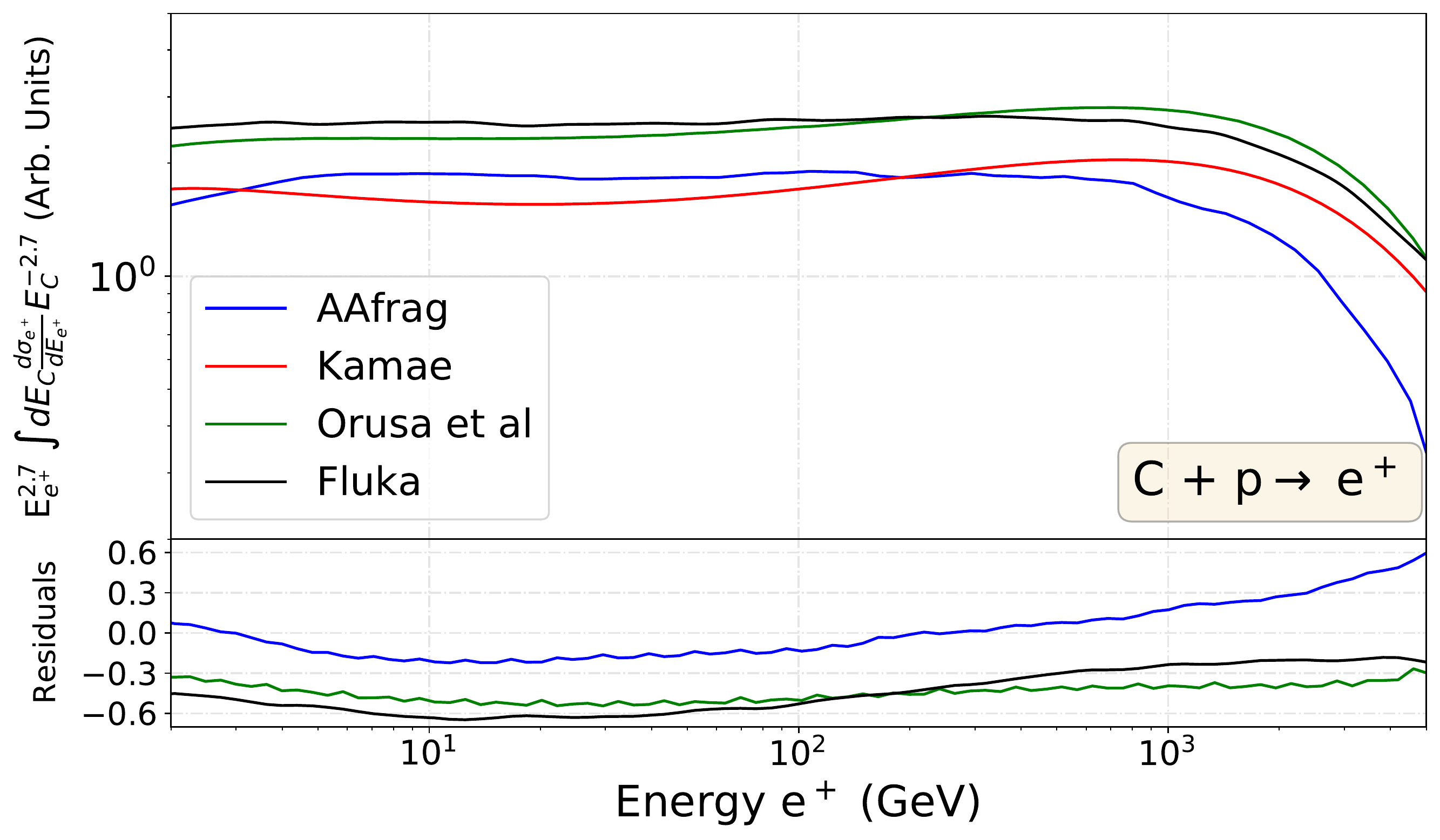}
    \includegraphics[width=0.49\textwidth]{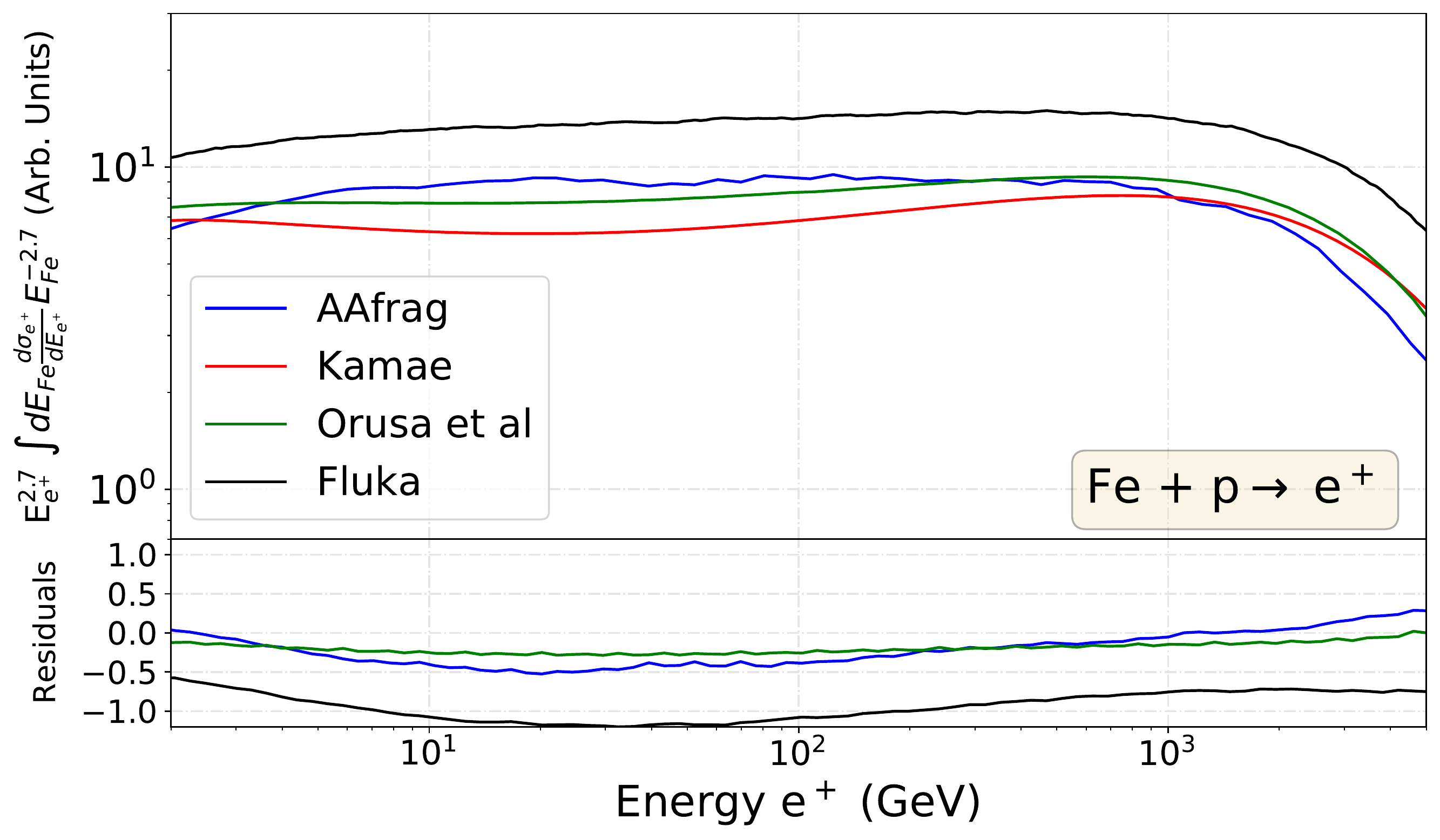}
    \caption{Total inclusive positron cross sections, weighted with a $E^{-2.7}$ function resembling the spectrum of CRs, for {\tt FLUKA} for the p-p (top-left panel), p-$^4$He (top-right panel), $^4$He-p (middle-left panel), $^4$He-$^4$He (middle-right panel), $^{12}$C-p (bottom-left) and $^{56}$Fe-p (bottom-right panel) channels, from $2$~GeV to $10$~TeV, compared to those from Refs.~\cite{Orusa, Kachelriess, Kamae_2006}. Residuals with respect to the Kamae cross sections are shown at the bottom of each panel.}
    \label{fig:SigmaTotpos}
\end{figure*}

\begin{figure*}[!th]
    \centering
    \includegraphics[width=0.49\textwidth]{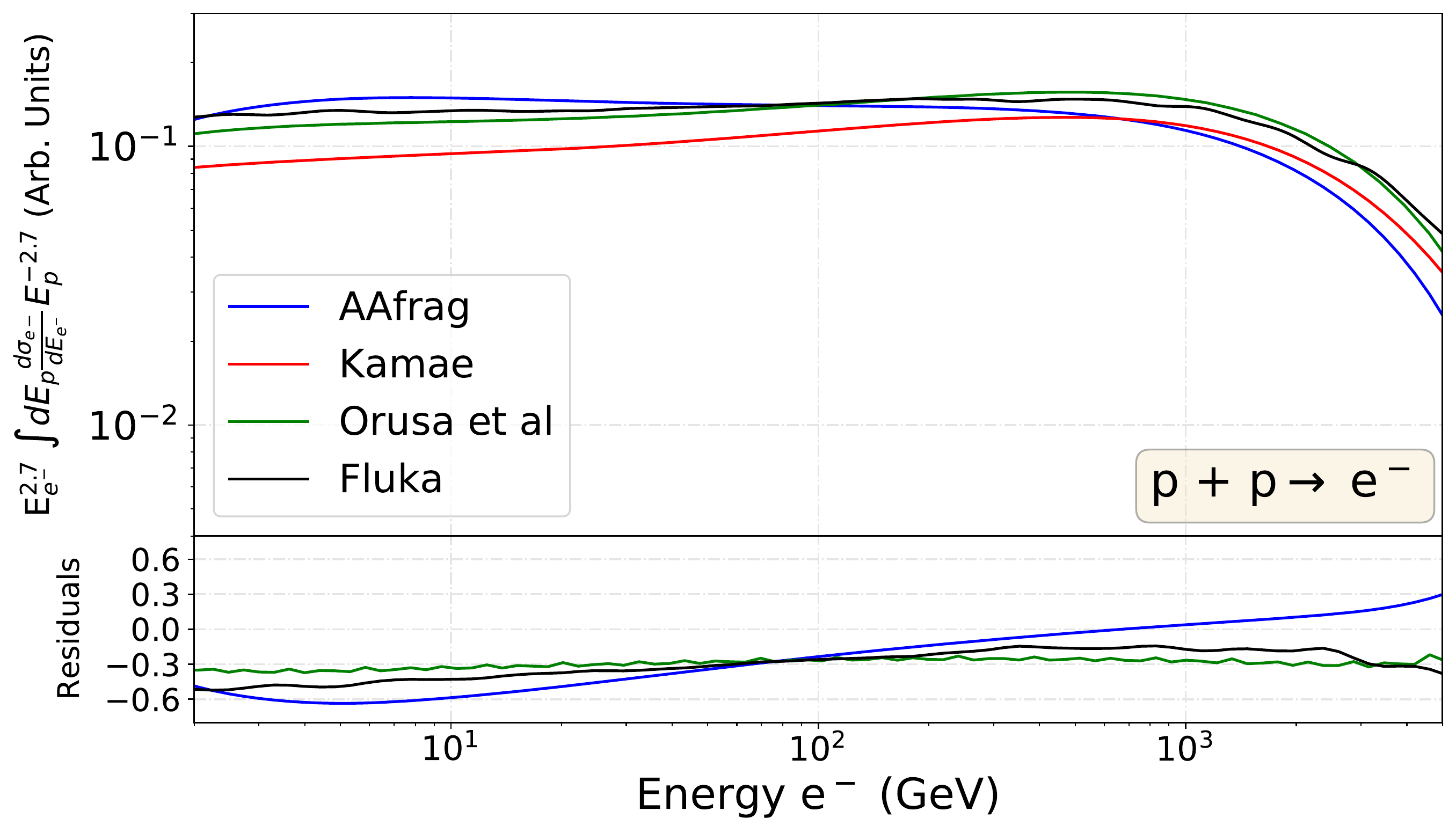}
    \includegraphics[width=0.49\textwidth]{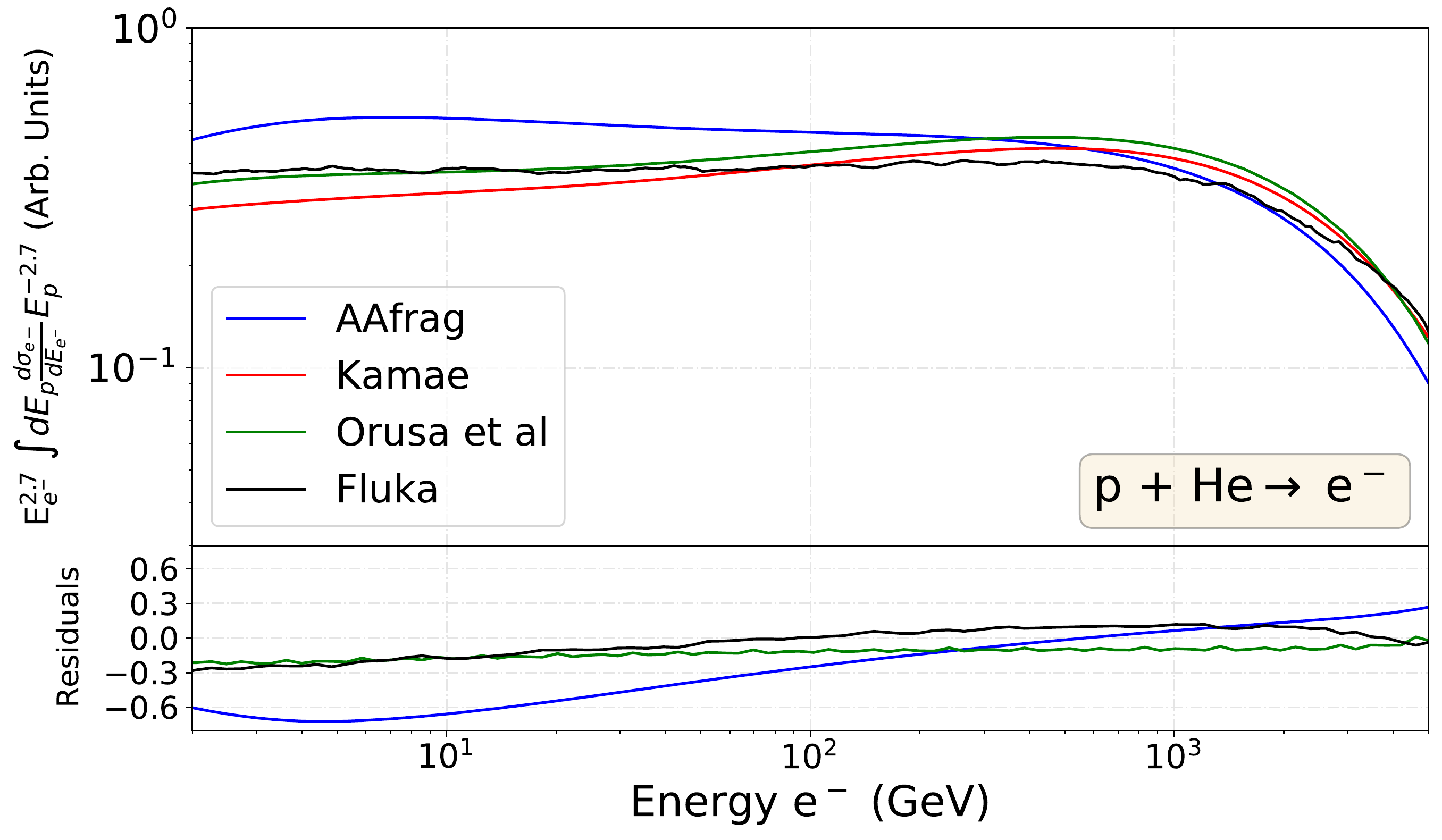}
    \includegraphics[width=0.49\textwidth]{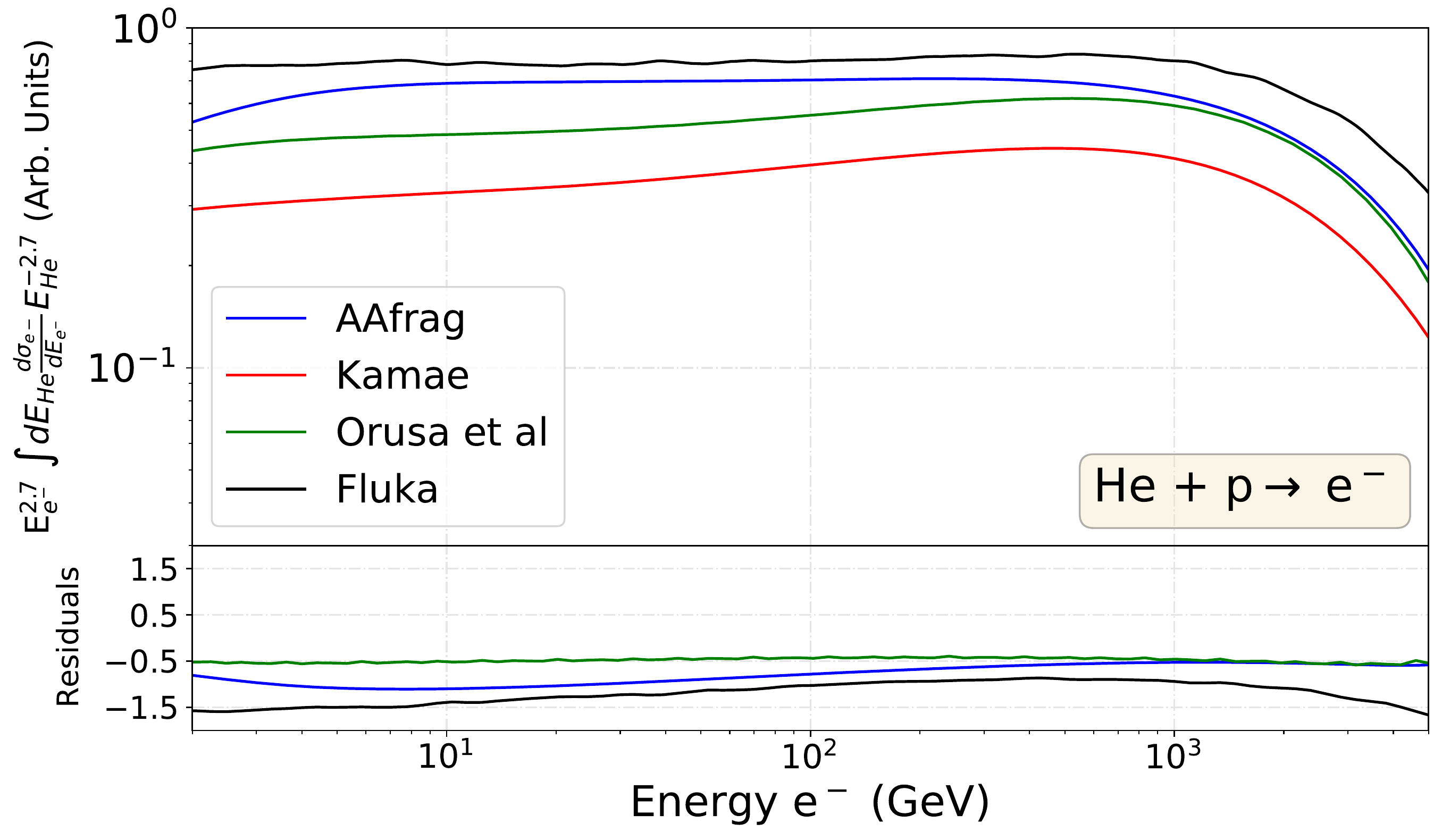}
    \includegraphics[width=0.49\textwidth]{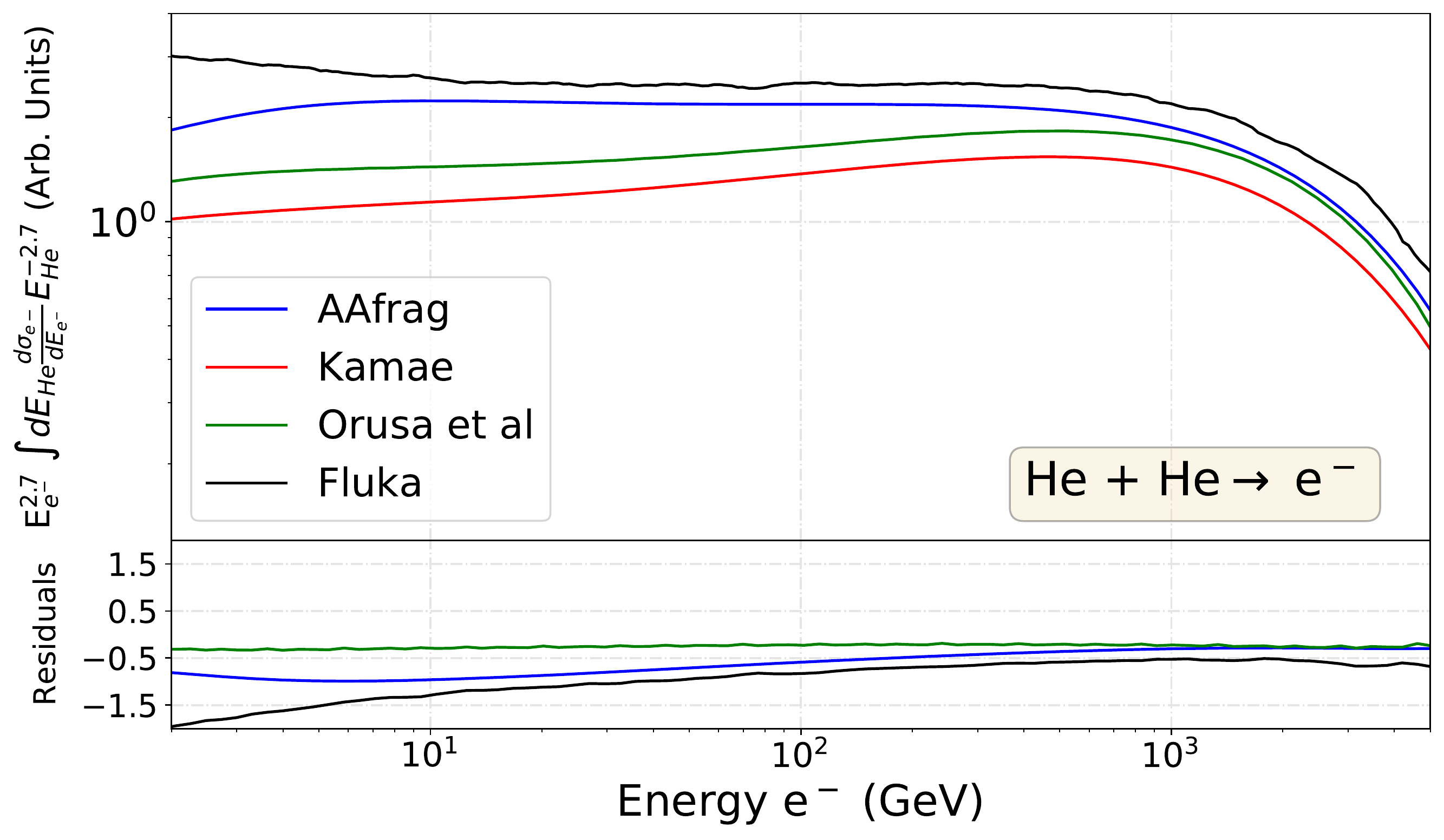}
    \includegraphics[width=0.49\textwidth]{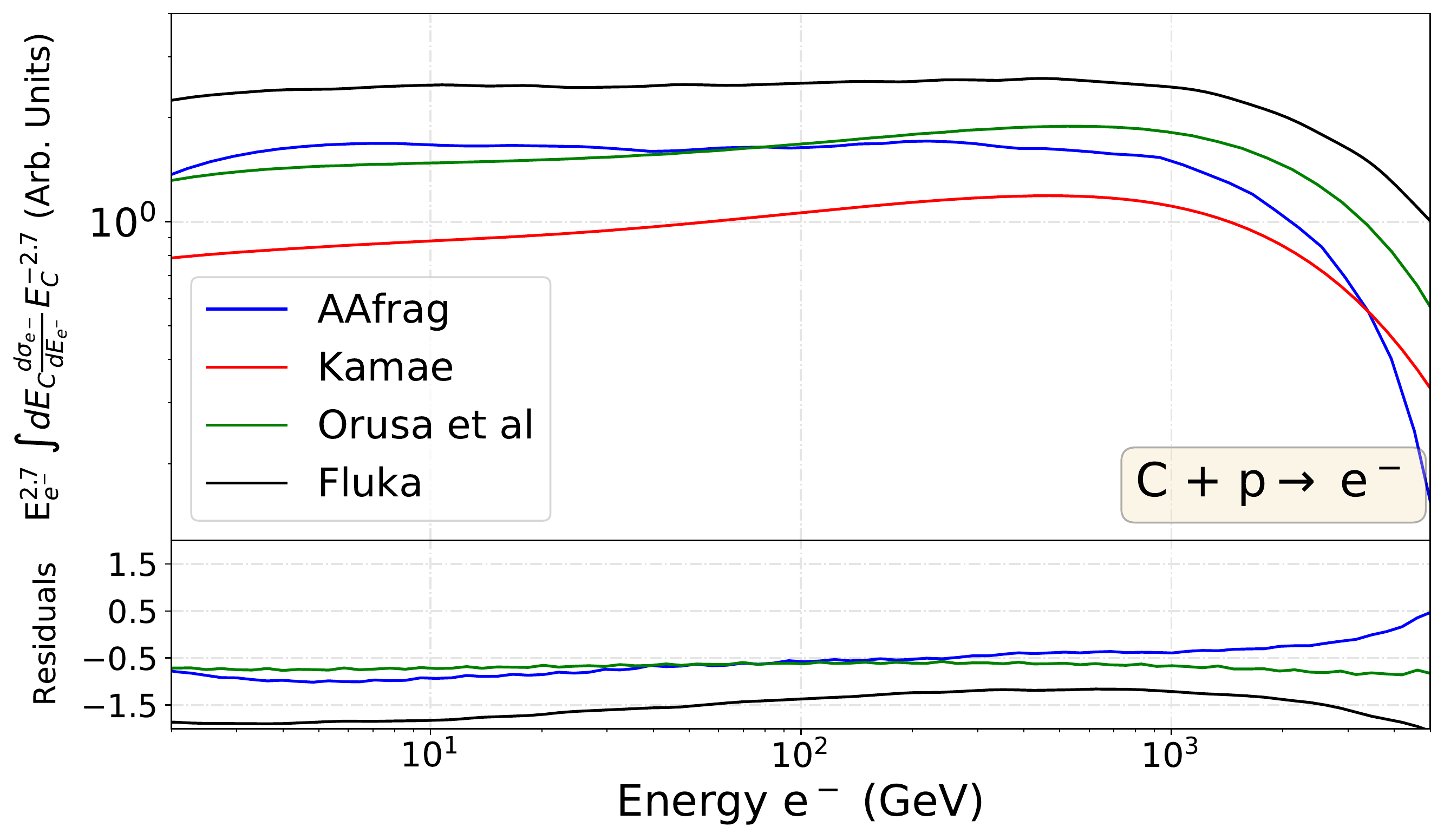}
    \includegraphics[width=0.49\textwidth]{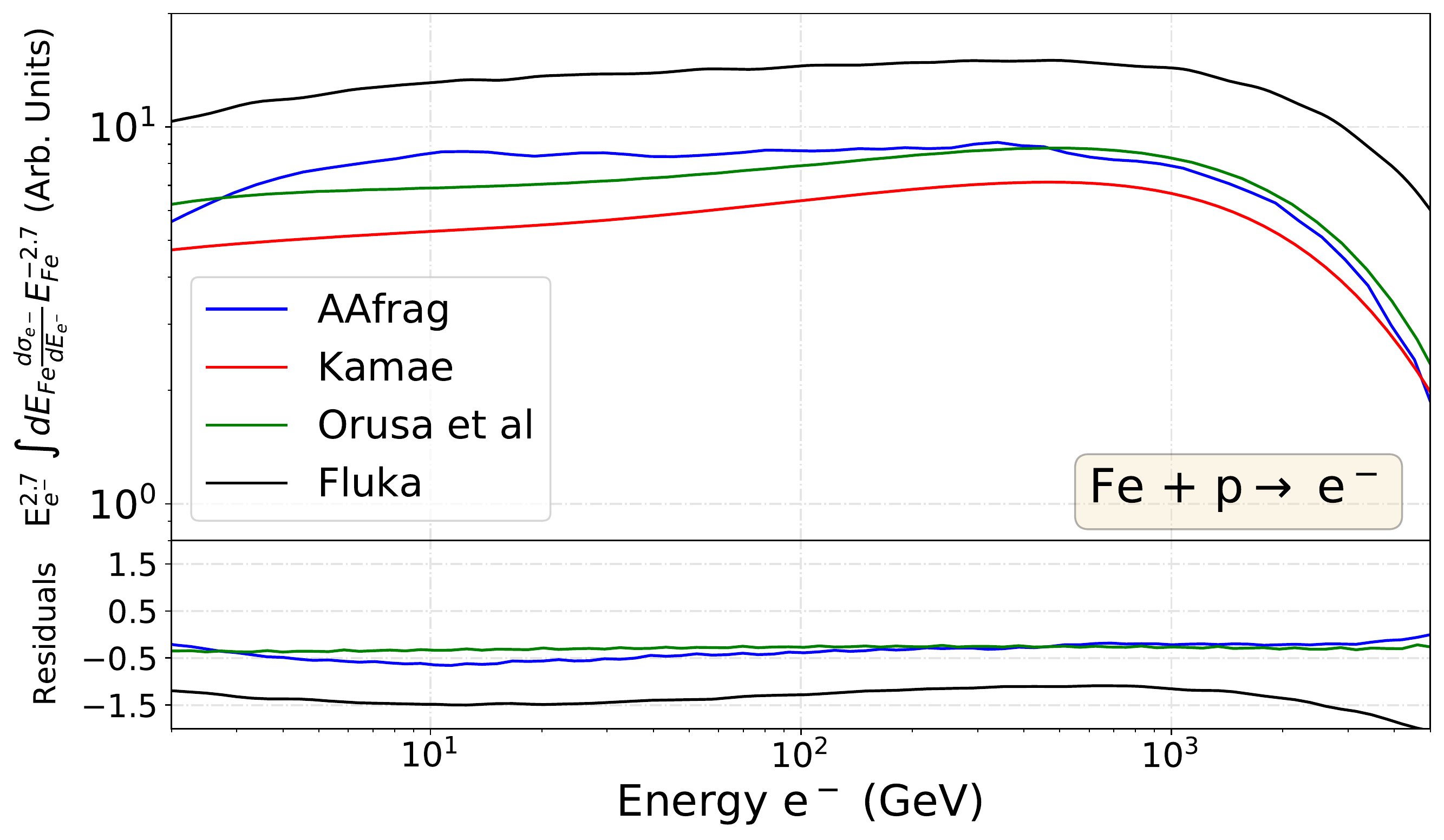}
    \caption{Similar to what is shown in Fig.~\ref{fig:SigmaTotpos}, but for the inclusive electron cross sections.}
    \label{fig:SigmaTotEl}
\end{figure*}

In Figure~\ref{fig:SigmaTotpos} (Fig.~\ref{fig:SigmaTotEl}) we show the total positron (electron) production cross sections obtained in our calculations for p-p (top-left panel), p-$^4$He (top-right panel), $^4$He-p (middle-left panel),  $^4$He-$^4$He (middle-right panel), $^{12}$C-p (bottom-left) and $^{56}$Fe-p (bottom-right panel) interaction channels, from $2$~GeV to $10$~TeV. These cross sections are calculated as the differential cross sections integrated over the energy of the projectile times a function of projectile energy that resembles the energy spectrum of CR particles ($E^{-2.7}$), similarly to the source term calculation, taking the upper limit for the projectile energy to be $35$~TeV/n.
In the figure, we compare the FLUKA inclusive cross sections with other popular cross-section datasets.
The \emph{Kamae} cross sections~\cite{Kamae_2006}, which constitute the most widely used dataset for CR leptons, are shown as a red line. These cross sections were obtained through parameterisations to non-diffractive, diffractive and resonance-excitation p-p interactions that relied in different formulae and Monte Carlo generators (see also Ref.~\cite{Kamae2005}). It is usual to extrapolate the p-p cross sections to the cross sections of interaction of heavier nuclei by using a simple scaling relation. Therefore, for the panels showing cross sections of interactions of heavier nuclei, the p-p Kamae cross sections are scaled by a factor $(A_{proj} \times A_{target})^{0.9}$, where $A_{proj}$ and $A_{target}$ are the mass number of the projectile and target, respectively. We employ the factor $0.9$, instead of the classical $2/3$ approximation, since it is favored from the analyses of cross sections measurements of heavy elements~\cite{Orusa}. 
Then, the AAfrag cross sections~\cite{AAfrag, Kachelriess}, which rely on a recently improved version of the QGSJET-II-04m Monte Carlo generator~\cite{QGSJET} to compute the yield of various secondary particles from p-p, p-nucleus and nucleus-nucleus interactions, are shown as a blue line. In particular, we are using the AAfrag202 software available at \url{https://sourceforge.net/projects/aafrag/files}.
Finally, the green lines correspond to the recent cross sections parameterisation obtained by \emph{Orusa et al.}~\cite{Orusa}, where the authors collected the most recent measurements~\cite{NA49, NA61SHINE, BRAHMS, PHENIX, CMS, LHC} of production of pions, kaons and $\Lambda$ resonances (see the channels that they included in their Fig.1) from p-p and nucleus-p interactions to obtain e$^{\pm}$ cross sections constraints from $1$~GeV to $\sim10$~TeV. Below $1$~GeV, where there is no available cross-section data for pion production, they extrapolate their parameterisation down to $10$~MeV.

The residuals are also shown with respect to the Kamae prediction (as (Kamae-model)/Kamae). The differences between the total cross sections evaluated from the different data-sets provide an estimation of the current uncertainty in the evaluation of the CR lepton cross sections for each channel.
For the positron cross sections, we observe the Kamae cross sections are systematically lower than the other cross sections for all channels, in the energy region below $\sim100$~GeV.
For the p-p channel, the inclusive cross sections of positron production from FLUKA, AAfrag and Orusa et al differ by no more than $5-10\%$ in the $5-100$ GeV region, exhibiting a very similar energy dependence. The Kamae cross sections are around a $20-30\%$ lower in this channel. In turn, for the He-p channel we observe that the AAfrag and Orusa et al cross sections are higher than those of Kamae by less $\sim15\%$ percent uo to $\sim200$~GeV, while the FLUKA cross sections are lower by $5-10\%$ in this energy range. For the channels with He as projectile, the discrepancies with respect to Kamae are of $40\%$ for the FLUKA cross sections in the He-p channel, again very similar to those from Orusa et al up to a few hundred Gev, but lower than $30\%$ in the He-He channel. 
In the case of electron inclusive cross sections there are more evident spectral discrepancies, being the Kamae prediction the lowest in every case below a few hundreds GeV. In both the p-p and p-He channels the FLUKA and Orusa et al cross sections are quite similar, being around a $30\%$ higher than Kamae. For channels with He as projectile the difference with respect to Kamae is of at least a $50\%$, with AAfrag and FLUKA showing very similar cross sections. The Orusa et al cross sections are more than a $20\%$ lower than FLUKA below $100$~GeV.
Then, as expected, we note more discrepancies in the channels involving heavier nuclei, finding differences of more than $\sim60\%$ for positrons and more than $100\%$ for electrons cross sections in those channels with He as projectile. 
For these channels, the FLUKA cross sections are higher than those from the other data-sets channels, being up to a factor of $2$ larger than the others for the Fe-p channel.

We have also cross-checked the differential inclusive cross sections at different projectile energies (see appendix~\ref{sec:AppendixB}), finding them to be usually in agreement within $20-40\%$ for the p-p channel, as shown in the left panel of Fig.~\ref{fig:He_contrib}.
For completeness, we report the differential cross sections of positron (electron) production as function of the projectile energy and the produced positron (electron) energy in Figs.~\ref{fig:2Delepos} and~\ref{fig:2Delepos_1}, for collisions of p, He, C, N and O with p and He.

\vskip 0.1in

\textbf{Contribution of heavy CRs}

The contribution of channels involving nuclei heavier than He has been commonly neglected. Here we report the contribution from the different interaction channels to the total source term of positron production and show how the fraction of heavy nuclei vary for different cross-section data-sets studied in this work.

\begin{figure}[t]
\includegraphics[width=0.48\textwidth] {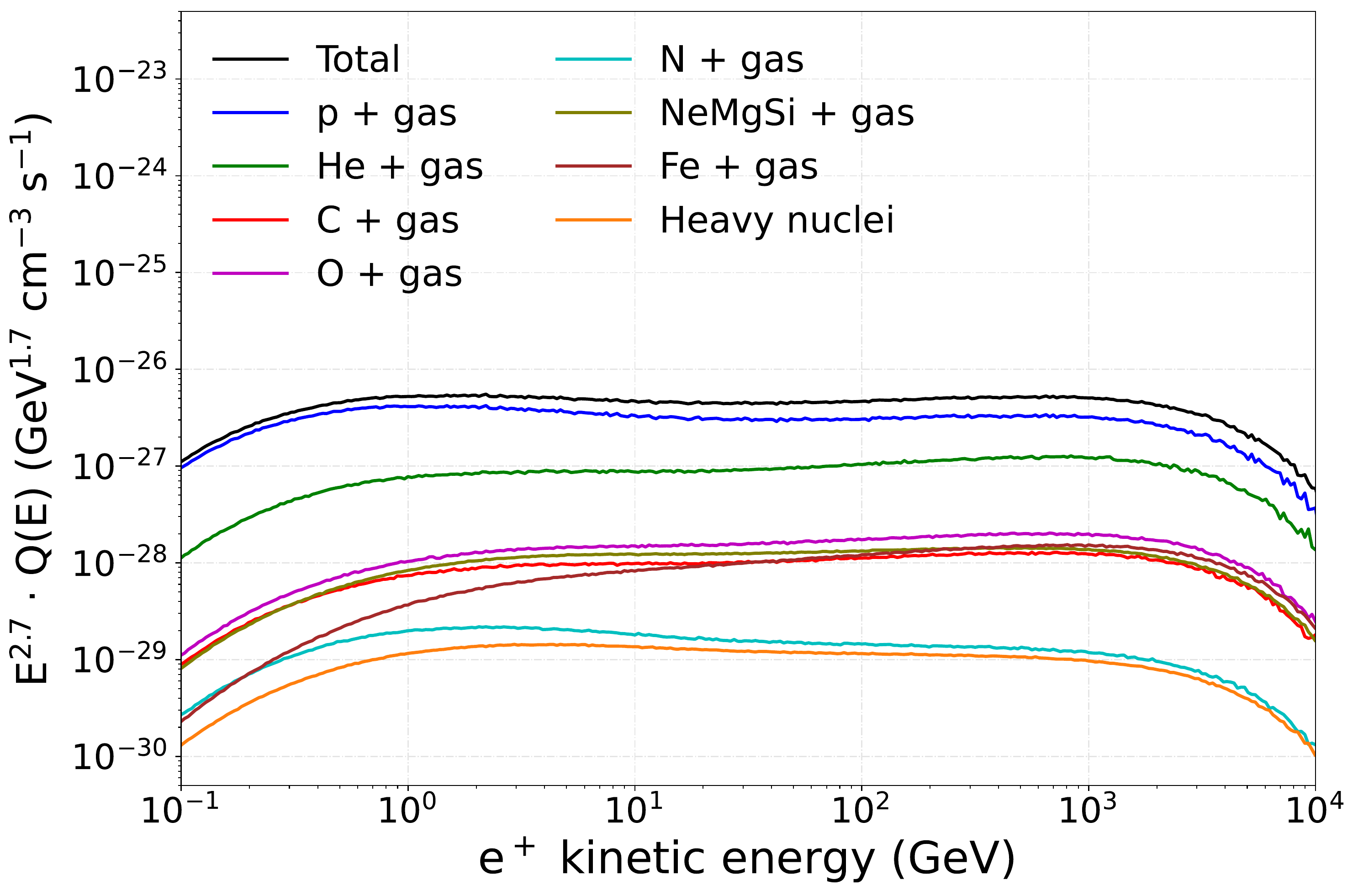} \hspace{0.3cm}
\includegraphics[width=0.48\textwidth] {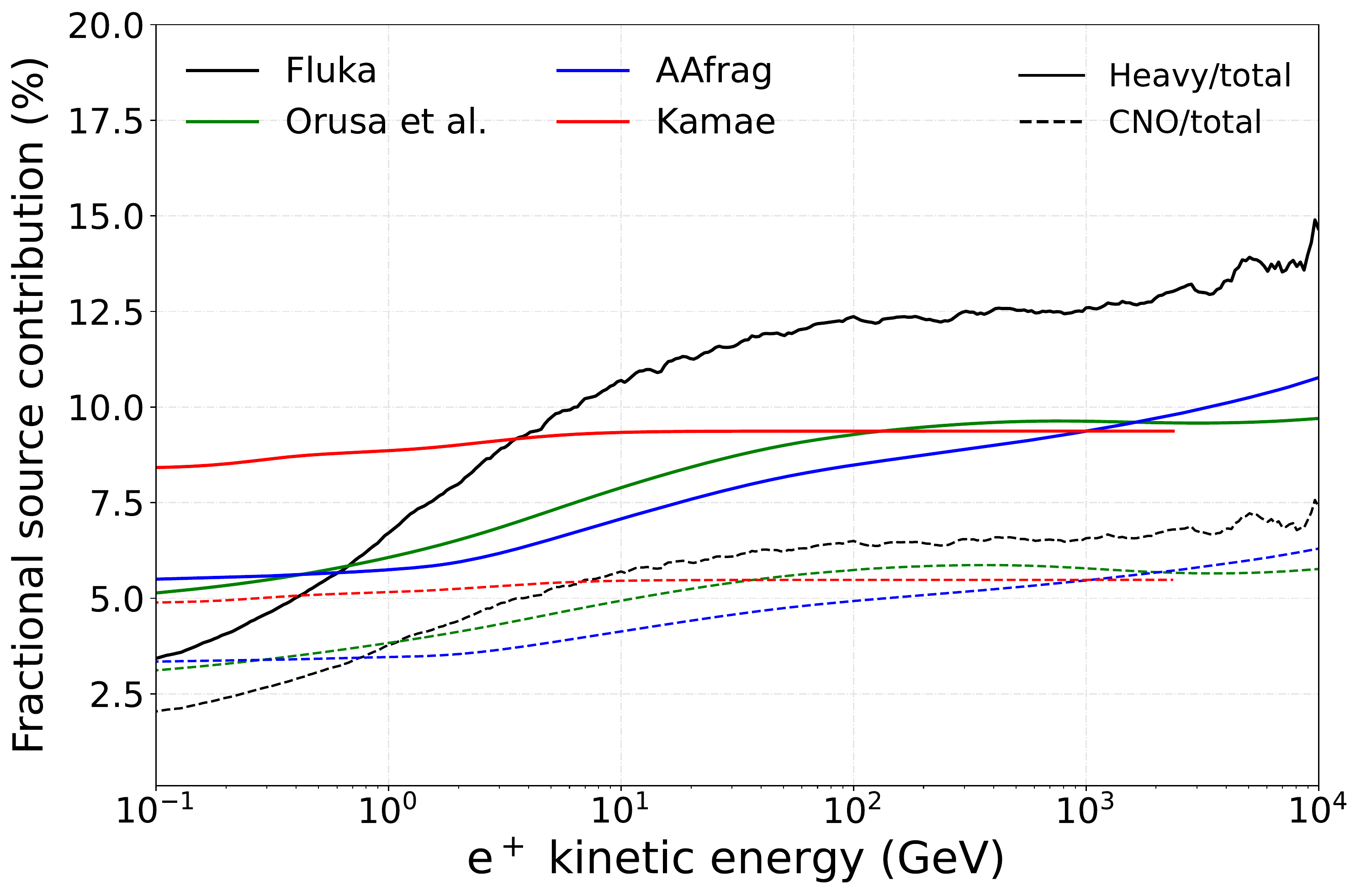}
\caption{\textbf{Left}: Source term contributions for positrons derived from {\tt FLUKA} for the main CR projectiles interacting with ISM gas (composed by $90\%$ of H and $10\%$ of He), assuming a gas density of $1$cm$^{-3}$. The \textit{Heavy nuclei} contribution is the sum of the source terms of CRs that with atomic numbers between 14 (Si) and 56 (Fe). 
\textbf{Right}: Fraction of the total source term for the sum of the C, N and O contributions (dashed lines) and for the sum of all nuclei heavier than He (solid lines) for the different data-sets discussed in this work compared to the {\tt FLUKA} predictions. For Kamae, AAfrag and Orusa et al, in those interaction channels where the cross sections are not available, the p-p cross sections have been scaled following the $A^{0.9}$ relation.}
\label{fig:SourceRatio}
\end{figure}

First, in the left panel of Figure~\ref{fig:SourceRatio}, we show the source term contributions for positrons derived from {\tt FLUKA} for the main CR projectiles interacting with ISM gas (composed by $90\%$ of H and $10\%$ of He), assuming a density of 1 cm$^{-3}$. In the legend, \textit{Heavy nuclei} stands for those CRs that with atomic numbers between Si and Fe (from S to Mn). The CR spectra used are obtained from fits to AMS-02 data~\cite{AMS_gen, Aguilar:2015ooa, Aguilar:2018keu, Aguilar:2020ohx, aguilar2017observation, aguilar2018observation, aguilar2019towards, AMS_NNaAl, AMS_Fe, AMS_Pos2019, AMS_F} in~\cite{delaTorreLuque:2022vhm}. 
Remarkably, we find that the Fe source term is at the level of the C source term around $20$~GeV and slightly higher above. This is due to the fact that the inclusive cross sections of positron production from Fe interactions are much higher than those from  lighter elements, compensating at high energies the lower flux of CR Fe. 
Similarly, the sum of the contributions from the NeMgSi group appears to be slightly higher than that from C.

In the right panel of Fig.~\ref{fig:SourceRatio}, we show the fraction of the total source term for the sum of the C, N and O contributions (dashed lines) and for the sum of all nuclei heavier than He (solid lines), for the different data-sets discussed above in comparison to the {\tt FLUKA} predictions. The Kamae cross sections are scaled for interactions of heavy nuclei, following the $A^{0.9}$ relation commented above. Given that the Orusa et al cross sections are available for nuclei up to O, the same scaling has been applied for nuclei heavier than O. This has been done also for the AAfrag cross sections, for the channels of O, Ne, Mg and Si. 
As we see in this figure, these data-sets agree fairly well in the expected contribution of the CNO group to the positron flux, above a few GeV. However, for the contribution of all heavy nuclei we find more sizeable differences, with a fractional contribution ranging from $\sim8\%$, for the Orusa et al data-set, to $\sim 11\%$, obtained for the FLUKA cross sections at $\sim10$~GeV. 
Note that in the evaluation of the CR fluxes, we do not inject any of the sub-iron nuclei, as Mn, Cr, V, Ti or Ca,
meaning that the contribution fraction of heavier nuclei to the positron flux can reasonably reach the level of $15\%$ above tens of GeV. In the right panel of Fig.~\ref{fig:He_contrib} we compare the contribution of He to the total source term for the different cross sections, obtaining a very good agreement between the FLUKA and Orusa et al. cross sections (around $17\%$ at $10$~GeV) and a very significant discrepancy with the AAfrag predictions.

\label{sec:Uncerts}

With the goal of building updated estimations of the expected positron flux at Earth, we implemented the FLUKA cross sections, as well as the other cross-section data-sets discussed above, into the {\tt DRAGON2} code\footnote{\label{note1} The version of the code is publicly available at the URL \url{https://github.com/tospines/Customised-DRAGON-versions/tree/main/Custom_DRAGON2_v2-Antinuclei}}~\cite{DRAGON2-1, DRAGON2-2}. In this way, we are able to produce detailed calculations of the propagation of these particles in the Galaxy, for different distributions of sources, gas density, magnetic field intensity, and the photon fields, considering all isotopes (till $Z=56$, i.e. Fe) contributing to the production of positrons and electrons. These estimations do not consider photo-pion production (reactions of the type $p + \gamma \rightarrow \pi^+ + n$), since their low cross sections make them contribute negligibly to the positron spectrum. Similarly, 
beta-decaying nuclei, such as $^{26}$Al, are not expected to produce significant amounts of local positrons (unless nearby sources of these elements are present).

In these computations, the injection of primary CRs is evaluated from a fit to a broken power-law (with break at $8$~GV) of the most recent AMS-02 data~\cite{AMS_gen, Aguilar:2015ooa, Aguilar:2018keu, Aguilar:2020ohx, aguilar2017observation, aguilar2018observation, aguilar2019towards, AMS_NNaAl, AMS_Fe, AMS_Pos2019, AMS_F}. These experimental data are taken from \url{https://lpsc.in2p3.fr/crdb/}~\cite{Maurin_db1, Maurin_db2} and \url{https://tools.ssdc.asi.it/CosmicRays/} \cite{ssdc}. We inject $^{1}$H, $^{4}$He, $^{12}$C, $^{16}$O, $^{20}$Ne, $^{24}$Mg and $^{28}$Si, $^{14}$N, $^{23}$Na $^{27}$Al and $^{56}$Fe, which are fit independently. All the details of the procedure where presented in Refs.~\cite{delaTorreLuque:2022vhm} and~\cite{Luque_MCMC}.
Then, the spatial diffusion coefficient is parameterised as \begin{equation}
 D_x = D_0 \beta^{\eta}\frac{\left(R/R_0 \right)^{\delta}}{\left[1 + \left(R/R_b\right)^{\Delta \delta / s}\right]^s} ,
\label{eq:DiffCoeff}
\end{equation}
where $R_0$, the rigidity at which the diffusion coefficient is normalized, is set to $4\units{GV}$.
For the parameters of Equation~\ref{eq:DiffCoeff}, we use the values $\Delta\delta = 0.14$, $R_b = 312 \units{GV}$ and $s = 0.040$, as explained in~\cite{Luque_MCMC} and derived in Ref.~\cite{genolini2017indications}. 
The diffusion coefficient in momentum space is related to the spatial diffusion coefficient following~\cite{osborne1987cosmic, seo1994stochastic, Strong_1998}:
\begin{equation}
    D_{pp} (R) = \frac{4}{3}\frac{1}{\delta' (4 - \delta'^2)(4 - \delta')} \frac{V_A^2 p^2}{D_x(R)} , 
    \label{dpp} 
\end{equation}
where $p$ is the momentum of the particle, $V_A$ is the Alfv\'en velocity ($V_A$ is defined as $\frac{B}{\sqrt{\mu_0 \rho}}$, being $\rho$ the plasma mass density, $\mu_0$ the magnetic permeability of the vacuum and $B$ the intensity of the magnetic field in the plasma) and $\delta'=\delta$ below $R_b$ and $\delta'=\delta-\Delta \delta$ above it.

Moreover, energy losses may impact very significantly the spectrum of CR positrons at GeV energies. In this energy regime, the relevant processes are synchrotron and inverse-Compton (IC) losses~\cite{Strong_1998}, which are proportional to the energy of the particle squared and the energy density of the magnetic and photon fields, respectively. For energies above the GeV these processes dominate over bremsstrahlung and represent the main mechanism of energy loss for CR electrons and positrons. This leads to the concept of electron and positron \emph{propagation horizon}, given that high energy particles are not able to propagate long distances from their source: the higher the energy where we measure the flux of these particles, the more this flux is dominated by the emission from nearby sources. This is the reason why the positron flux at high energies is expected to be dominated by the emission of local PWN. 

We use, as default configuration, the magnetic field distribution derived by Ref.~\cite{Pshirkov_2011} (PT11), with a normalization of the disk's and halo's magnetic field intensity to B$_{0,disk} = 2$~$\mu$G and B$_{0,halo} = 4$~$\mu$G, respectively, and the turbulent field to B$_{0,turb} = 7.5$~$\mu$G, as found in Ref~\cite{DiBernardo_2013} from the study of synchrotron radiation. The energy density distribution of the radiation fields has been taken from Ref.~\cite{porter2008inverse}. 
Then, we set as default, a smooth, cylindrically symmetric gas distribution derived from Ref.~\cite{Bronfman1988} and incorporated to the numerical implementation of the GALPROP code by Refs.~\cite{Strong_ICRC, Strong_1998}, that we refer to as \textit{2D Galaxy model}. Finally, for the distribution of supernova remnants in the Galaxy, we use the Ferriere et al. distribution~\cite{ferriere2007spatial} as described in Ref.~\cite{DRAGON2-1}.

In the next sub-sections we explore different modifications of the main ingredients affecting our estimations of the local positron flux in order to show the uncertainties derived.

\subsection{Cross sections and diffusion parameters}
\label{sec:Uncs-XS_Diff}

With the setup discussed above, we compute the propagated e$^{+}$ spectrum for each cross-section data-set for two different sets of propagation parameters. This approach allows us to illustrate the uncertainties originating from the cross section and from the transport of CRs.
These propagation parameters have been found from the fit with a Markov chain Monte Carlo analysis of the AMS-02 B/C spectrum and a combined fit to the CR flux ratios of B, Be and Li with C and O (also from AMS-02), reported in Ref~\cite{delaTorreLuque:2022vhm}. 

The B/C ratio has been traditionally used to infer the transport parameters. The best-fit parameters found from the fit to the B/C ratio are: $D_0=6.5$~cm$^2$/s, $\eta=-0.53$, $\delta=0.43$, $V_A=25.8$~km/s. These are well compatible with other analyses performed before~\cite{Luque_MCMC, derome2019fitting}, but due to cross sections uncertainties of B production, they show a slightly larger ($\sim10-20\%$) $D_0$ value with respect to those derived from the GALPROP or DRAGON2 cross sections~\cite{Luque:2021joz, Evoli:2019wwu, GALPROPXS, GALPROPXS1}.
Thanks to the precise data from AMS-02, the combination of different secondary CRs have been recently employed to mitigate the impact of nuclear cross sections uncertainties in the determination of the propagation parameters~\cite{Korsmeier:2021brc, Weinrich_combined, Luque:2021joz}. The parameters obtained in the combined fit performed in our companion work are: $D_0=9.8$~cm$^2$/s, $\eta=-0.81$, $\delta=0.36$, $V_A=39.7$~km/s. These propagation parameters allowed a simultaneous reproduction of not only all ratios of B, Be and Li, but also of the ratio $^3$He/$^4$He.
The halo height value found from a fit the the $^{10}$Be ratios was $7.5$~kpc, which was found to be compatible to other recent estimations~\cite{Weinrich_halo, CarmeloBeB}.
As we see, the two analyses yielded very different propagation parameters, and this allows us to understand the impact of the transport uncertainties in our predictions for the local positron flux. In particular, the parameter that shows the larger impact on the positron flux is the Alfv\'en velocity, which is around $V_A=25$~km/s for the B/C fit, but around $V_A=40$~km/s for the combined fit. The probability distribution function obtained in these fits, for each parameter, is compared as a corner plot in Fig.~\ref{fig:Corner_plot}.
We remark that the set-up employed to derive the propagation parameters, including reacceleration and a low-energy break in the injection spectrum of primary CRs,  has been recently demonstrated by Ref.~\cite{dimauro2023novel} to provide the best $\chi^2$  value from the fit to CR data.

\begin{figure}[t]
\includegraphics[width=0.48\textwidth] {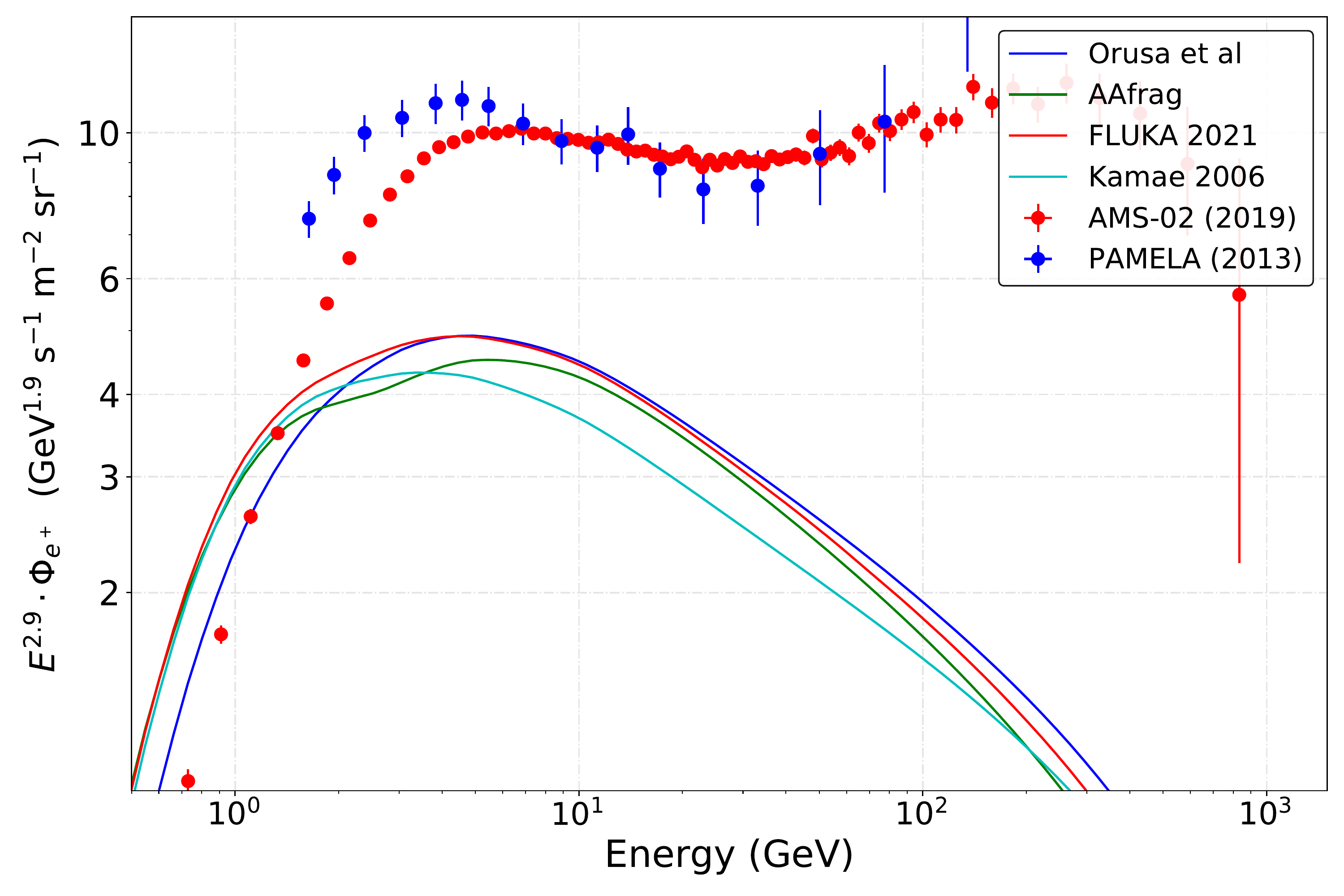}
\includegraphics[width=0.48\textwidth] {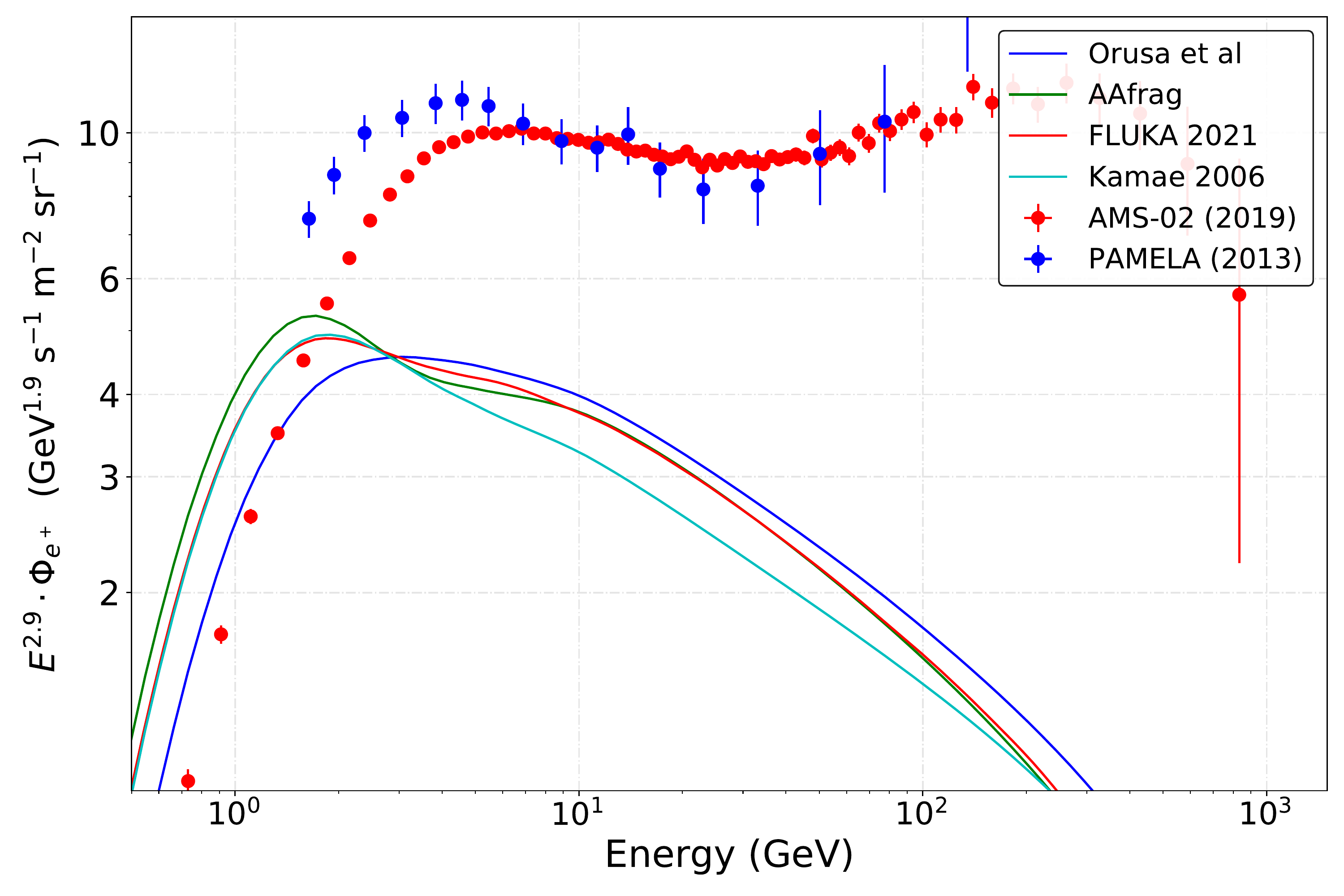}
\caption{
Local positron spectrum evaluated using the FLUKA, Kamae, Orusa et al. and AAfrag cross sections with the propagation parameters obtained in the fit to AMS-02 data of the B/C ratio (left panel) and the parameters obtained in the combined fit of the flux ratios of B, Be and Li to C and O  (right panel), from the analysis in Ref.~\cite{delaTorreLuque:2022vhm}. These propagation parameters were derived from a halo height value $H=7.5$~kpc. In these spectra  we account for the production of positrons from all isotopes up to Fe.
}
\label{fig:XScomp}
\end{figure} 

In Figure~\ref{fig:XScomp} we report  the estimated local positron flux for the FLUKA, Kamae, Orusa et al. and AAfrag cross sections with the propagation parameters found in the B/C fit (left panel) and in the combined fit (right panel).
On one hand, we see that the positron fluxes predicted with different cross sections may differ up to a $30\%$, being the Orusa et al. prediction the highest and Kamae the lowest, above $2$~GeV.  
We note that the predicted FLUKA spectra are compatible with those from Orusa et al. within a $10\%$ factor from $\sim1$~GeV to a few hundred GeV; however, below that energy, the predictions from the Orusa et al. cross sections falls more rapidly than the others.
On the other hand, we appreciate that the differences in propagation parameters do not lead to significant changes in the positron spectrum above $2-3$~GeV, where the effect of reacceleration is lower and the positron spectrum mostly depends on the spectra of H and He (fitted to the AMS-02 data, as shown in Fig.12 of Ref.~\cite{delaTorreLuque:2022vhm}) and the cross sections. The most significant difference between the spectra derived from the two propagation setups is the peak due to reacceleration below $3$~GeV, that appears very clear in the spectra computed using the parameters derived from the combined fit, and is caused by reacceleration of lower energy positrons. 
Finally, we remark that, for all the cross sections, the predicted positron flux inferred from the B/C propagation parameters is maximum at around $4-5$~GeV, coinciding with the energy at which we see a sort of bump in the AMS-02 data. This seems common for all predictions with small $V_A$ and likely indicates that the low energy part of the positron spectrum must be dominated by secondary production. 

It is important to note that one can be tempted to constrain the Alfv\'en velocity from low energy positron measurements, however there are two further sources of uncertainty affecting very significantly the estimations of the local positron flux at these energies that do not allow us to do it: the halo size and the Solar modulation, that are explored in the next subsections. Nevertheless, the fact that the positron spectrum that we observe peaks at $\sim4$~GeV and falls down quickly may indicate that reacceleration does not play a major role in the positron spectrum.
In these calculations, we estimate the effect Solar modulation by using the simplistic force-field approximation~\cite{forcefield}, setting the Fisk potential to $\phi=0.59$~GV, consistent with the rest of CRs for the period of collection of the positron data and the measurements from the NEWK neutron monitor experiment\footnote{Available at \url{http://www01.nmdb.eu/station/newk/} \cite{MaurinCRDB, MAurinCRDB2}} (see~\cite{ghelfi2016non,ghelfi2017neutron}).

\subsection{Galactic halo height}

The determination of the height of the Galactic halo, H, is a key issue for the study of astroparticles in the Galaxy, affecting not only our predictions on the spectra of CRs in the Galaxy, but also the current indirect dark matter searches with CRs (see, e.g. the recent study by Ref.~\cite{John_2021}). 
Although there have been studies trying to constrain the halo height from radio observations of synchrotron~\cite{bringmann2012radio}, X-ray or gamma-ray~\cite{biswas2018constraining} emissions, the observable that seemed to be more promising to constrain H was the measurement of the flux of the unstable isotope $^{10}$Be and its ratios. However, the CR community has recently emphasised the impact of cross sections uncertainties for production of Be isotopes in our evaluations~\cite{Maurin_Be10}, only allowing us to constrain H to be between $\sim3$ and $12$~kpc~\cite{Korsmeier:2021bkw, Thesis}. In addition, other systematic uncertainties further worsen these constraints. 

\begin{figure}[t]
\includegraphics[width=0.48\textwidth] {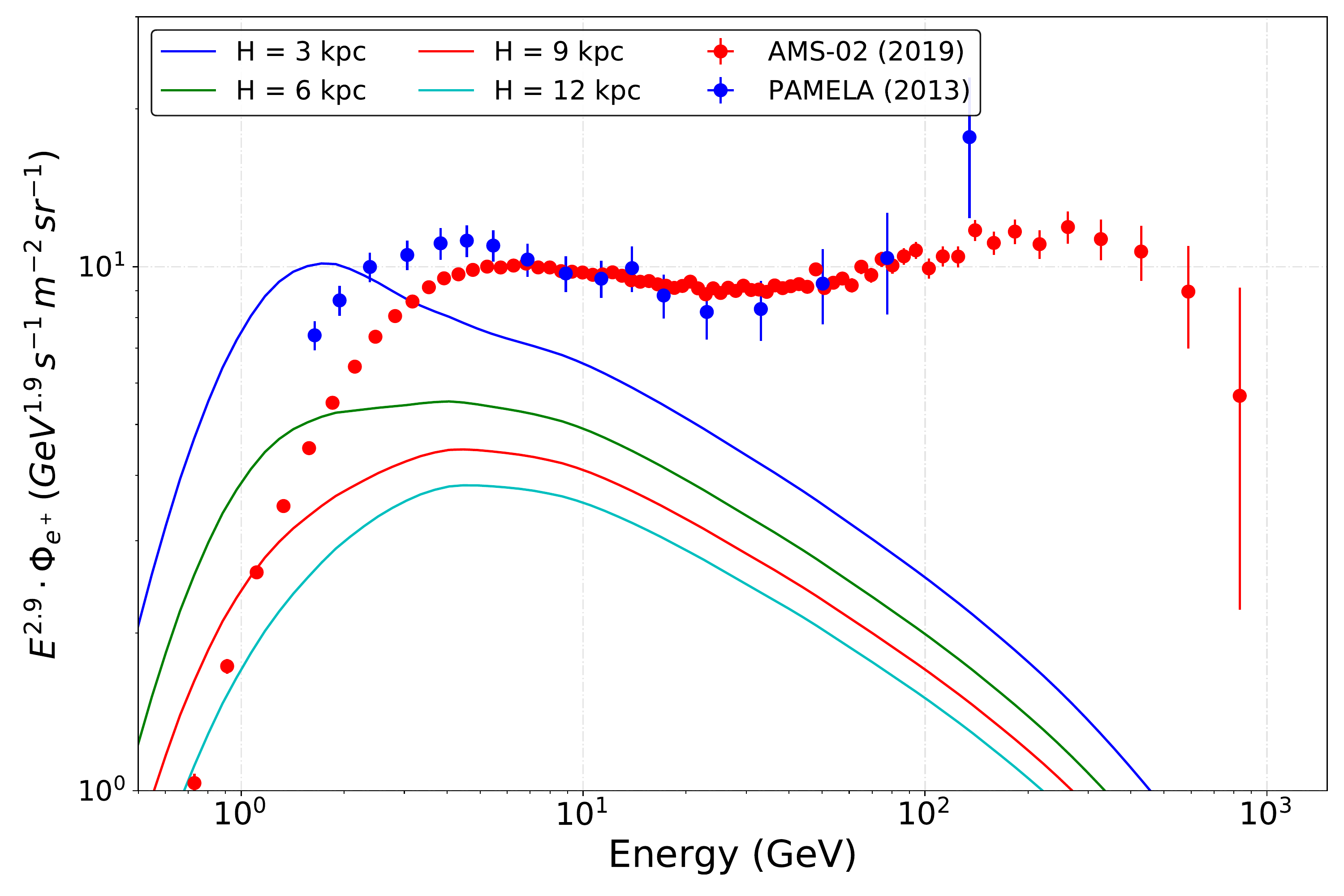}
\includegraphics[width=0.48\textwidth] {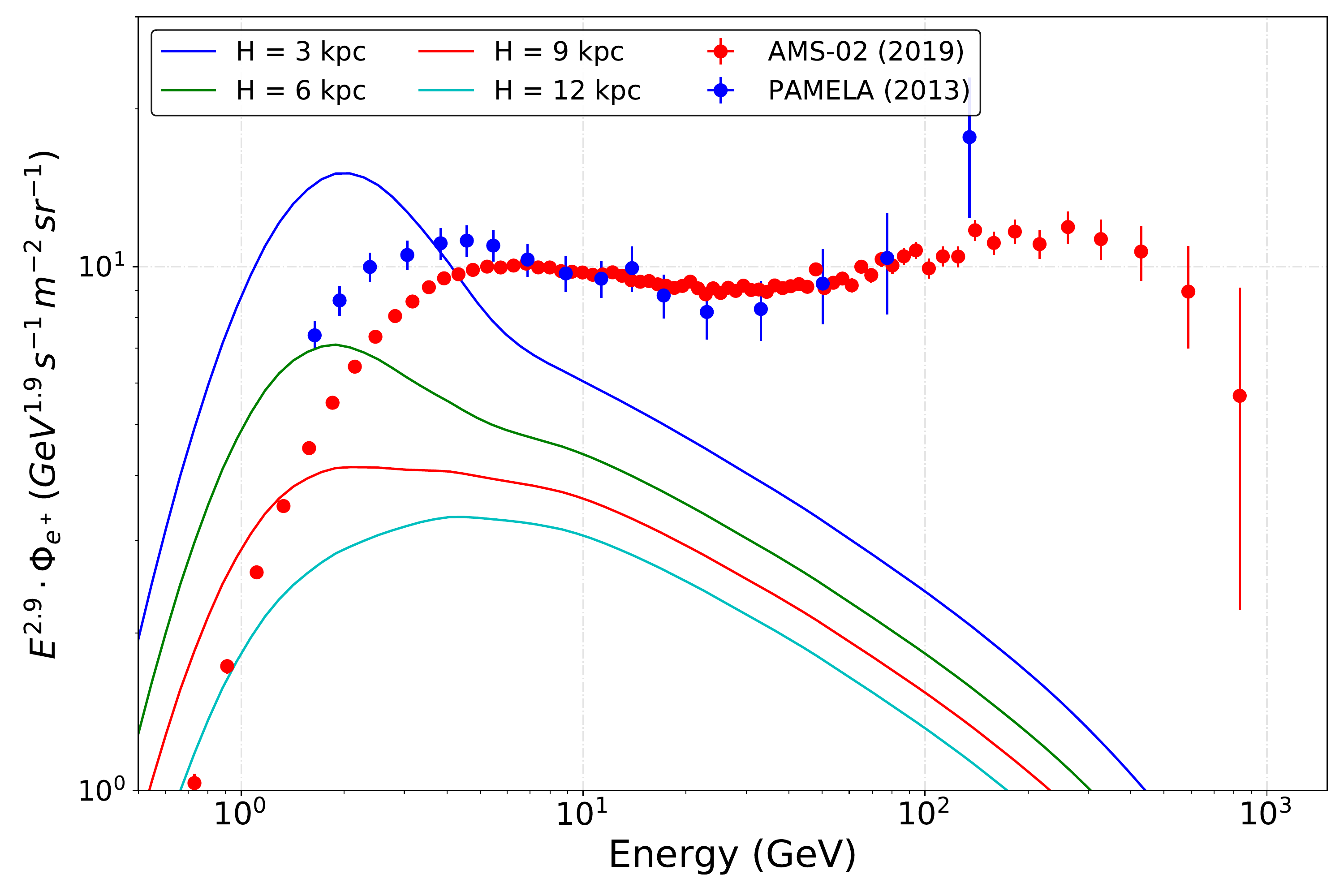}
\caption{Local positron spectrum (using the FLUKA cross sections) derived for different values of the halo height, with the propagation parameters obtained from the fit to B/C data (left panel) and the combined fit (right panel). In every case, the normalization of the diffusion coefficient, $D_0$ is adjusted such that the B/C spectrum reproduces the observations from AMS-02. These calculations consider the production of positrons from all isotopes up to Fe.}
\label{fig:Hcomp}
\end{figure}

The effect of uncertainties in the determination of the halo height is particularly relevant for positrons, since radiative (IC and synchrotron) energy loss rates are of the same order of magnitude as the reciprocal of the diffusion time~\cite{Lavalle_Pos}. This can be understood in the following way: the halo size determines the volume of propagation, and the smaller the volume of propagation the more densely secondary particles are produced (and vice versa) for the same amount of grammage traversed by CRs (where the grammage is proportional to the ratio $D/H$). This explains what we see from Fig.~\ref{fig:Hcomp}, where we report the predicted local positron spectrum (using the FLUKA cross sections) for different values of the halo height, from $3$ to $12$~kpc. In every case, the normalization of the diffusion coefficient, $D_0$ is adjusted such that the B/C spectrum reproduces the observations from AMS-02. Moreover, we remind the reader that in these calculations we include the production of positrons from all isotopes up to Fe.

A few important points should be highlighted from these comparisons: 
\begin{itemize}
    \item  As we expect, a smaller halo height leads to a much larger local flux of positrons, scaling roughly as a normalization factor above a few GeV. 
    \item Again here, we do not observe significant differences above $2-3$~GeV between the predictions from the different diffusion parameters used, since, at these energies, the positron fluxes mostly depend on the flux of primary CRs (mainly H and He). At lower energies, the effect of reacceleration are very clear.
    \item The peak in the positron spectra at low energies, caused by the reacceleration of low energy positrons, is more pronounced for low values of H. This is due to the fact that reacceleration is proportional to $1/D_x$ and the diffusion coefficient is smaller for low halo heights to match the measurements of secondary-to-primary flux ratios.
    \item There have been a few studies imposing constraints on the halo height using positrons (see, e.g. Refs~\cite{Weinrich_halo, Lavalle_Pos}). However, we emphasize that, given the uncertainties in the determination of 
    $V_A$ and the effect of Solar modulation, this is not an easy task. In fact, we note here that, with the basic setup employed here, the only estimations not surpassing the data are those for $H>12$~kpc. Nevertheless, as we show below, this can be easily amended for reasonable changes of our default setup.
\end{itemize}

\subsection{Effect of Solar modulation}

The transport of charged particles in the Heliosphere is still poorly understood~\cite{Potgieter_2013}, essentially due to the variability of the Solar magnetic field, drift effects, non-linearities and the anisotropy involved in the process. Although advanced codes devoted to the computation of modulated CR fluxes have been developed~\cite{Vittino_2017, KAPPL2016386, Engelbrecht_2013, BOSCHINI20182859} showing successful results~\cite{Shalchi_2006, MaccionePRL, Boschini_2020}, the complexity of this process and the lack of measurements outside the Heliosphere limit our predictions and make us unable to produce robust estimations. 

Modulation uncertainties affect the spectra of CR particles below a few tens of GeV and the discrepancies in the predictions from different modelling approaches below $\sim2$~GeV are specially important, given the lack of data and the how the Solar winds are modelled. This affects quite considerably dark matter searches with CR antiparticles~\cite{Potgieter_antiCR, TimAp_modulation}, since there are systematic uncertainties that are commonly not accounted for. For instance, the force-field approximation does not consider the effect of Solar winds in CR particles, nor the charge-sign dependence of the modulation.

\begin{figure}[t]
\centering
\includegraphics[width=0.55\textwidth] {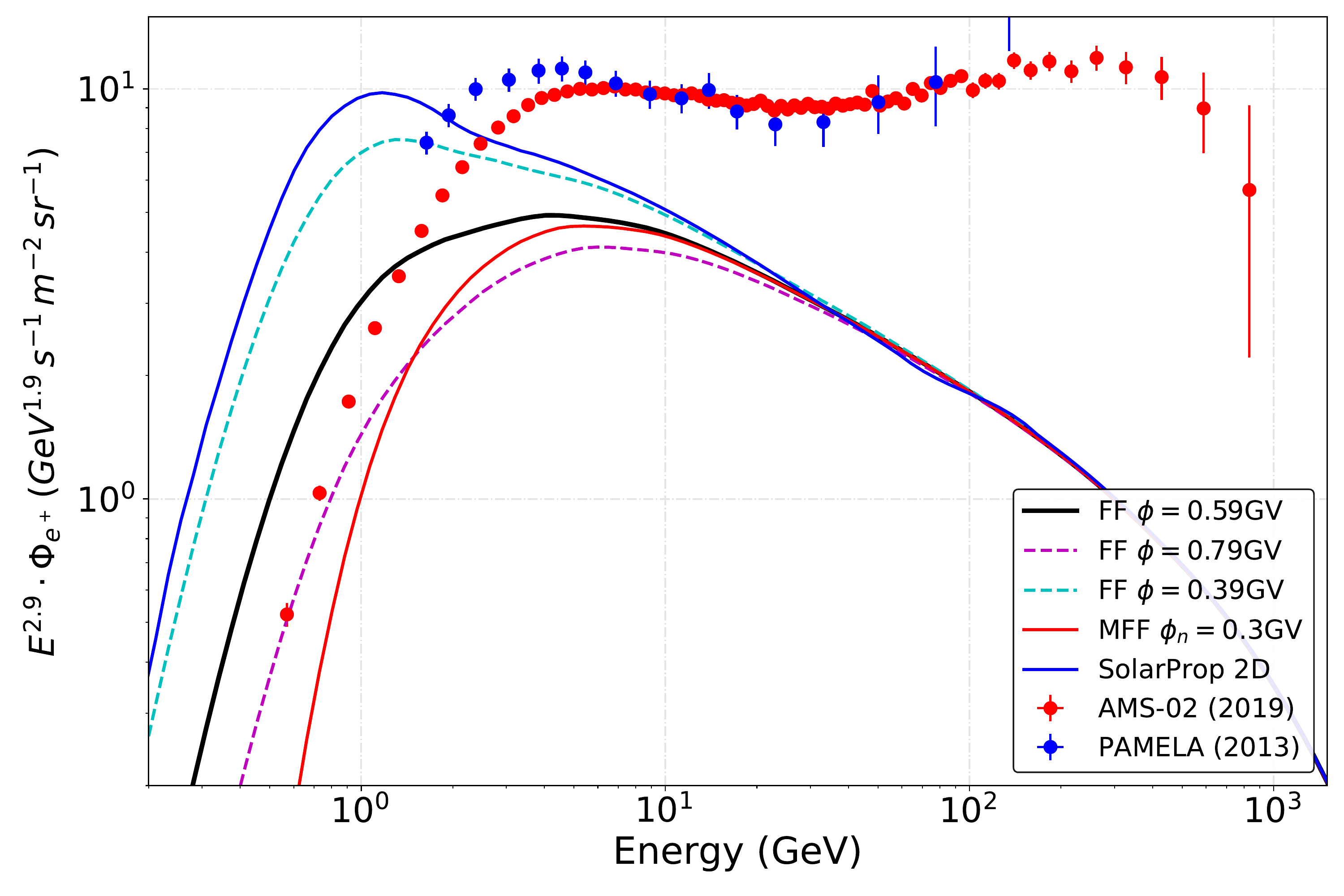}
\caption{Local positron spectra for different approaches to model the effect of Solar modulation. The black line represents the prediction using the force-field approximation with $\phi=0.59$~GV. The magenta and cyan dashed lines are obtained using $\phi=0.79$~GV and $\phi=0.39$~GV, respectively. The red line is obtained using the modified force-field model proposed by Ref.~\cite{Cholis:2015gna}, with $\phi=0.59$~GV and $\phi_n=0.2$~GV. The blue line is obtained using the SOLARPROP code with parameters adjusted to those fitting the AMS-02 proton using the \textit{Standard2D} model.}
\label{fig:Modul}
\end{figure}

Ref.~\cite{Weinrich_halo} found that uncertainties from Solar modulation are the dominant effect at $1$~GeV. They estimated a $50\%$ uncertainty in the positron flux by comparing the predicted flux when $\phi$ is varied by $\pm0.1$~GV, under the force-field approximation. However, this estimation of the uncertainty highly depends on the approach used to simulate the modulation and, mainly due to the effect of reacceleration, it depends on the propagation parameters used, at least below $2-3$~GeV.
In general, we observe that the higher the Fisk potential (i.e. the higher the shielding that the Heliosphere imposes to low-energy CRs), the less evident are the peaks related to reacceleration, and vice versa, as it can be seen from Fig.~\ref{fig:Modul}. 
 
In Fig.~\ref{fig:Modul}, we show the predicted local positron spectrum for different values of the Fisk potential in the force-field approximation, using the propagation parameters derived from the fit to B/C AMS-02 data. In particular, we show the spectra predicted using $\phi=0.59$~GV (black solid line), $\phi=0.39$~GV (cyan dashed line) and $\phi=0.79$~GV (magenta dashed line).
Then, we show the spectrum predicted by a modified version of the force-field approximation that was proposed to account for the charge-sign dependence of the modulation in Ref.~\cite{Cholis:2015gna}. This model parameterises the Fisk potential as $\phi(t) = \phi(t) + \phi_n(t)\frac{R_0}{R}$ where R denotes rigidity and $R_0=1$~GV. In this figure we show, as a red solid line, the predicted positron flux for $\phi=0.59$~GV and $\phi_n=0.2$~GV. As wee see, the predicted flux is similar to the one obtained from the traditional force-field approximation at $E>3$~GeV, while it gets drastically reduced for lower energies.
Finally, we show, as a blue solid line, the predicted spectrum using the SOLARPROP code~\cite{KAPPL2016386} with parameters adjusted to those fitting the AMS-02 proton flux with local interstellar spectrum obtained in Ref~\cite{delaTorreLuque:2022vhm}, using the \textit{Standard2D} model. Curiously, we observe that the flux obtained from this computation is quite different from the one obtained using the force-field approximation with $\phi=0.59$~GV.

\subsection{Gas density distribution}

\begin{figure}[t]
\includegraphics[width=0.40\textwidth] {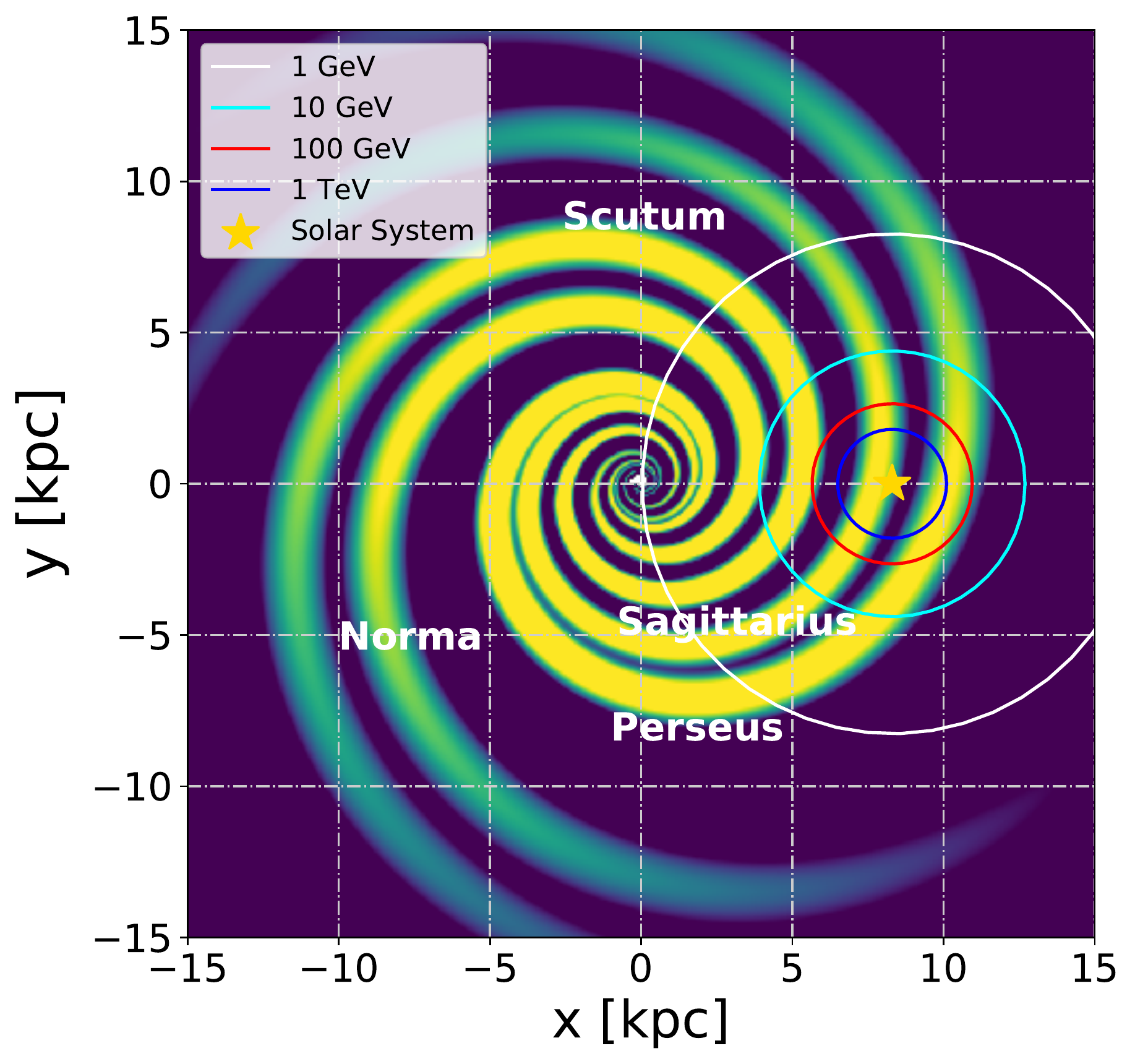}
\hspace{0.15cm}
\includegraphics[width=0.55\textwidth] {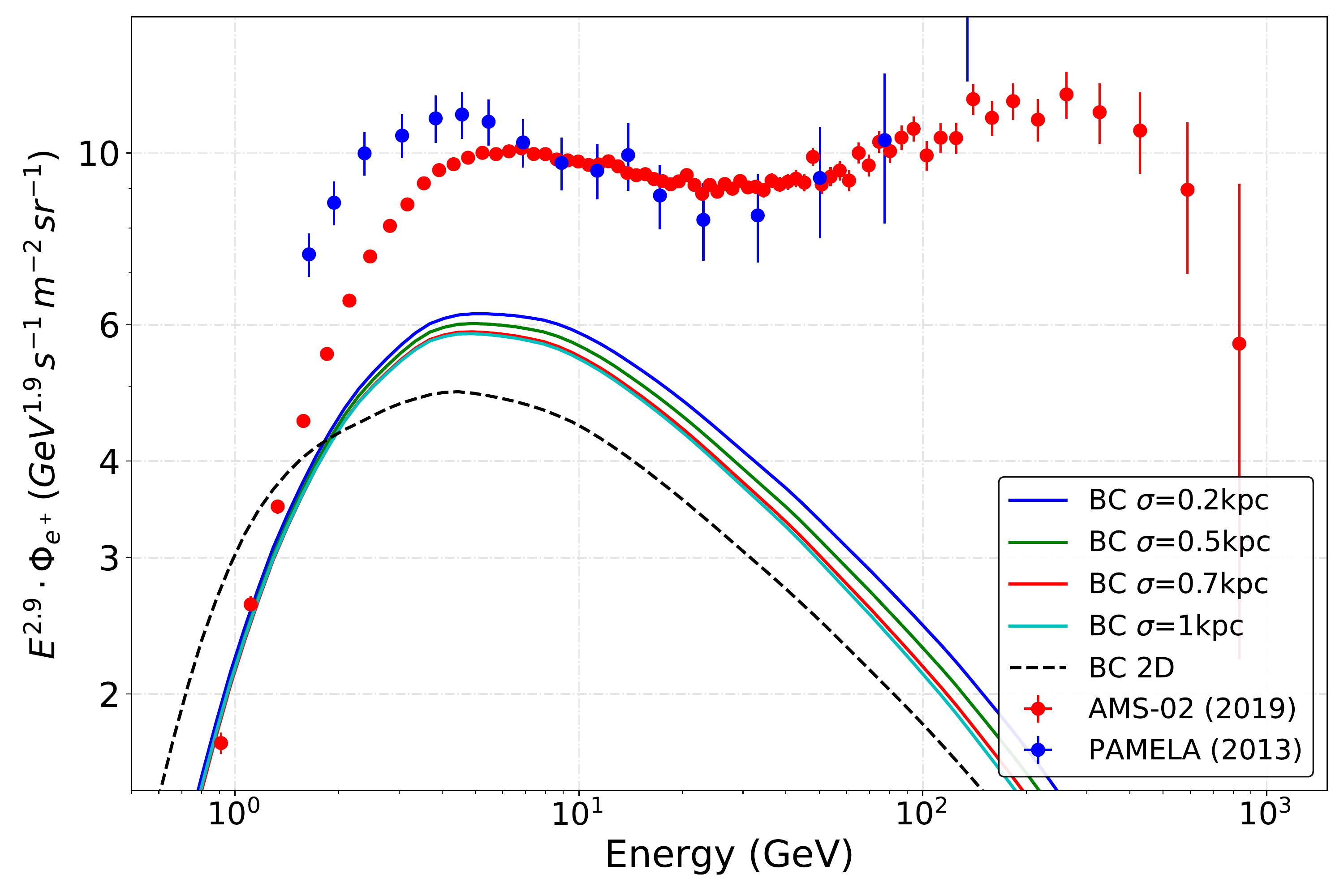}
\caption{\textbf{Left panel:} Galactic 3D gas density distribution adopted in these calculations~\cite{Steiman-Cameron_2010}, with the name of the four arms. The position of the Solar system, is represented by a star and the circles around the Earth show the average distance travelled by a positron with $10$, $100$, $10^3$~GeV, subject to IC and synchrotron energy losses, computed as detailed in the text. \textbf{Right panel:} Local positron spectra evaluated with the 3D Galactic model for different values of the arm width (from $\sigma=0.2$~kpc to $\sigma=1$~kpc), comparing to the predictions from our 2D model and for both the propagation parameters found from the B/C fit. In every case, the B/C spectrum is reproduced by adjusting $D_0$, as explained in the text.}
\label{fig:3D}
\end{figure} 

Given that IC and synchrotron energy losses do not allow electrons and positrons (above $\sim 1$GeV) to propagate for long distances, the local spectrum of these particles is mainly dominated by the flux produced in the vicinity of the Earth; the larger the energy, the more local is the origin from these particles, either if they are injected by sources or if they are produced from CR interactions. The transport parameters play an important role here, since the larger is the diffusion coefficient, the more these particles will travel for a certain time-scale. This simply means that we expect that the distribution of the interstellar gas density will affect
our estimations above the GeV, similarly, but in a different extent, to what happens for variations of the halo size. This has been also found by Ref.~\cite{Johannesson_2018}.

Therefore, we have tested how our predictions change with a more realistic gas density distribution that incorporates the spiral arm structure of the Galaxy, since this would imply that the local gas density would be significantly larger than the average one in the Galaxy. In particular, we adopt the Steiman-Cameron et al.~\cite{Steiman-Cameron_2010} gas distribution model for four spiral arms. In DRAGON2 the width of the arms are implemented as Gaussian functions around the center of the arm with a standard deviation (the width scale of the arms) that we can adjust. This distribution is convoluted with the Ferriere et al 2D (radial and vertical) distribution of the interstellar gas and is normalized such that the total amount of matter is the same as in the 2D model. The arm distribution is similarly implemented for the interstellar radiation fields (ISRF) energy density and the distribution of sources (see also Ref.~\cite{Daniele3D}). More details about the implementation of this distribution can be found in Ref.~\cite{DRAGON2-1}. 

In the left panel of Figure~\ref{fig:3D} we illustrate the Galactic 3D gas density distribution adopted, indicating the name of the four arms. Here, the position of the Solar system is represented with a star (the Earth lays in the Sagittarius arm). Moreover, with circles around the Earth, we show the average distance travelled by a positron with $10$, $100$, $10^3$~GeV, evaluated as $\langle r \rangle (E) = \sqrt{2 D(E) \tau_{loss}(E)}$, where $D$ is the diffusion coefficient obtained from the B/C fit and $\tau_{loss}$ is the energy loss time-scale considering only IC and synchrotron energy losses. For this estimation we calculate the energy loss rate as described in Refs.~\cite{DeLaTorreLuque:2022chz, Carmelo_ELoss}, considering a magnetic field of $3$~$\mu$G, the cosmic microwave background (CMB) field, infrared field, an optical field and three ultraviolet radiation fields, and the Klein-Nishina effect.
As we see from the figure, a $10$~GeV electron would travel, in average, $\sim3$~kpc, implying that most of the particles detected at Earth have been produced in the surroundings. Moreover, we expect that a $1$~GeV particle is still affected by the gas density distribution.

In the right panel of Fig.~\ref{fig:3D}, we report the local positron spectra evaluated with the 3D Galactic model for different values of the arm width ($\sigma$). An important point here is that the predicted local B/C ratio is also dependent on the gas density, so that we have adjusted the normalization of the diffusion coefficient, $D_0$ to reproduce the B/C ratio in every case. In particular, this implies that we are using a larger $D_0$ value~\footnote{Concretely, $D_0$ changes from $6.5\cdot10^{28}$cm$^{2}$s$^{-1}$ in the 2D case to $10.6\cdot10^{28}$cm$^{2}$s$^{-1}$ in the 3D case ($\sigma=0.5$)}, which causes that the effect of reacceleration is significantly mitigated for same values of $V_A$ (since $D_{pp} \propto 1/D_x$). In this figure, we show how different values of the arm width affect our estimated flux: a smaller arm width implies a larger gas density around the center of the arm, and, hence, a slightly larger local flux of positrons above the GeV. We report variations of the arm width from $\sigma=0.2$~kpc to $\sigma=1$~kpc for the propagation parameters obtained from the B/C fit. In comparison, we plot, as a dashed blue line, the spectrum evaluated from the 2D model for the parameters obtained from the B/C fit. We can appreciate differences between both setups that range from $15\%$ (for the case with $\sigma=1$~kpc) to $\sim30\%$ ($\sigma=0.2$~kpc). Additionally we note that the low energy feature resulting from reacceleration is not present in the 3D predictions, given that larger $D_0$ value employed. We have checked that in the case of no reacceleration present, the 2D and 3D predictions meet below $\sim1$~GeV.

Since different propagation parameters and other ingredients of this setup can modify the comparisons shown here, it is difficult to perform consistently a complete evaluation of the uncertainties related the gas distribution (see also Ref.~\cite{Johannesson_2018}).
However, these comparisons already demonstrate that neglecting the spiral arm density distribution can lead to an underestimation of the local positron spectrum above the GeV of up to $\sim40\%$.

Finally, we note here that the use of the spiral arm distribution also affects significantly the predicted spectrum of the positron anisotropy amplitude, as shown in Fig.~\ref{fig:Anisotropy} of the Appendix~\ref{sec:Anisotropy}. We estimate the dipole anisotropy following the diffusive-regime approximation~\cite{Ginz&Syr}: $\frac{3D(E)}{c} \frac{|\nabla f(E, \textbf{r})|}{f(E, \textbf{r})}$, where $f(E, \textbf{r})$ is the positron density and $D(E)$ is the spatial diffusion coefficient. We find that a different arm width may imply a difference of more than a factor of $2$ in the predicted dipole anisotropy amplitude. Also variations of the halo height yield differences in the predicted amplitude by a similar factor. However, these predictions still lay at least one order of magnitude below the upper limits recently reported by the AMS-02 collaboration~\cite{Velasco_2020}.

\subsection{Galactic magnetic field and synchrotron losses}

As discussed above, energy losses are a crucial factor in the evaluation of the electron and positron spectrum at Earth. While ionization and Coulomb losses dominate (have the shortest time-scale) up to the MeV energies, bremsstrahlung becomes dominant till a few GeV and IC and synchrotron losses take over above those energies. Bremsstrahlung losses depend on the gas density distribution, and the uncertainty associated to it is difficult to quantify. However, it is not expected that this uncertainty can imply a significant error in the estimation of the positron flux at $1$~GeV.

\begin{figure}[!th]
\begin{center}
\includegraphics[width=0.48\textwidth] {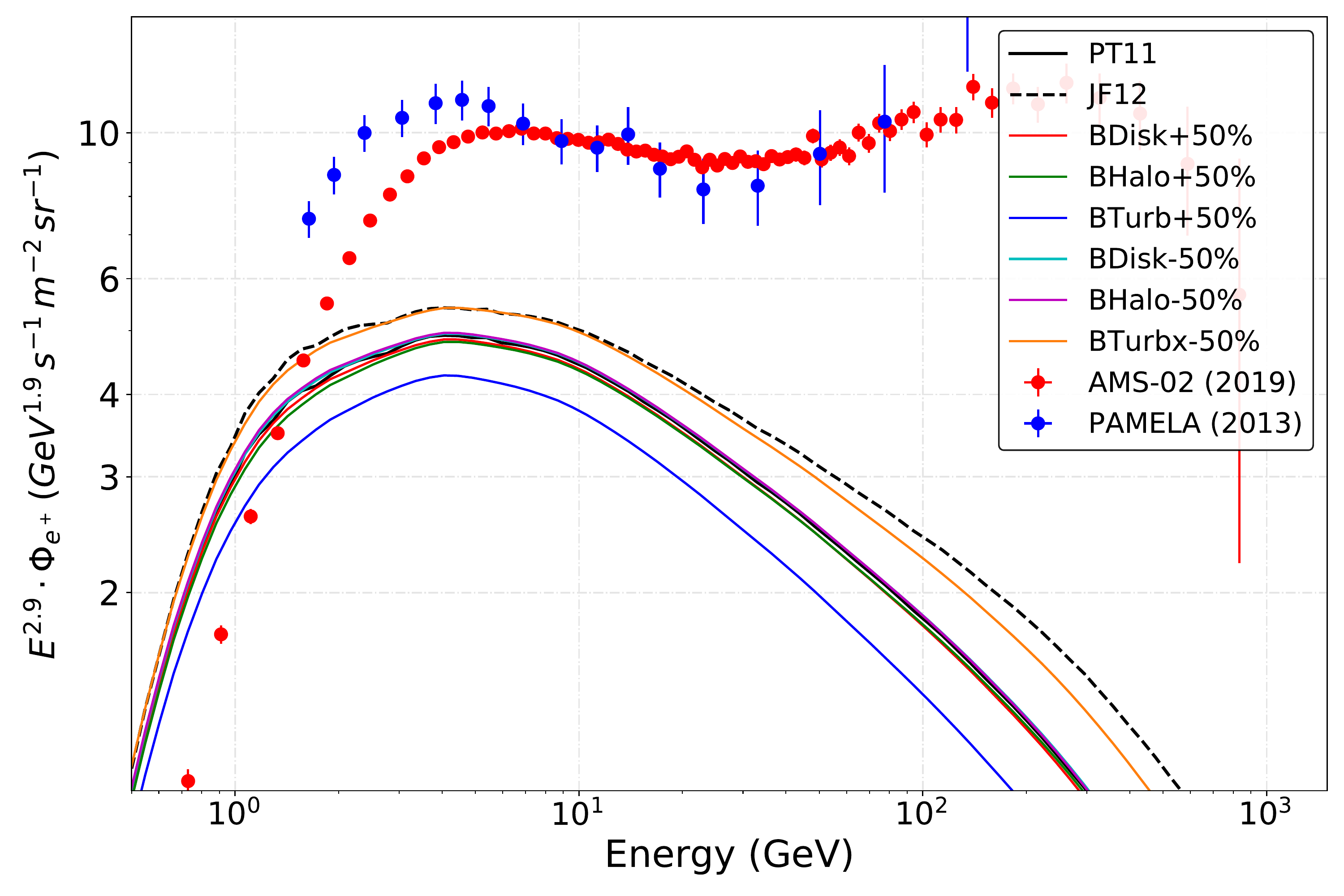}
\includegraphics[width=0.48\textwidth] {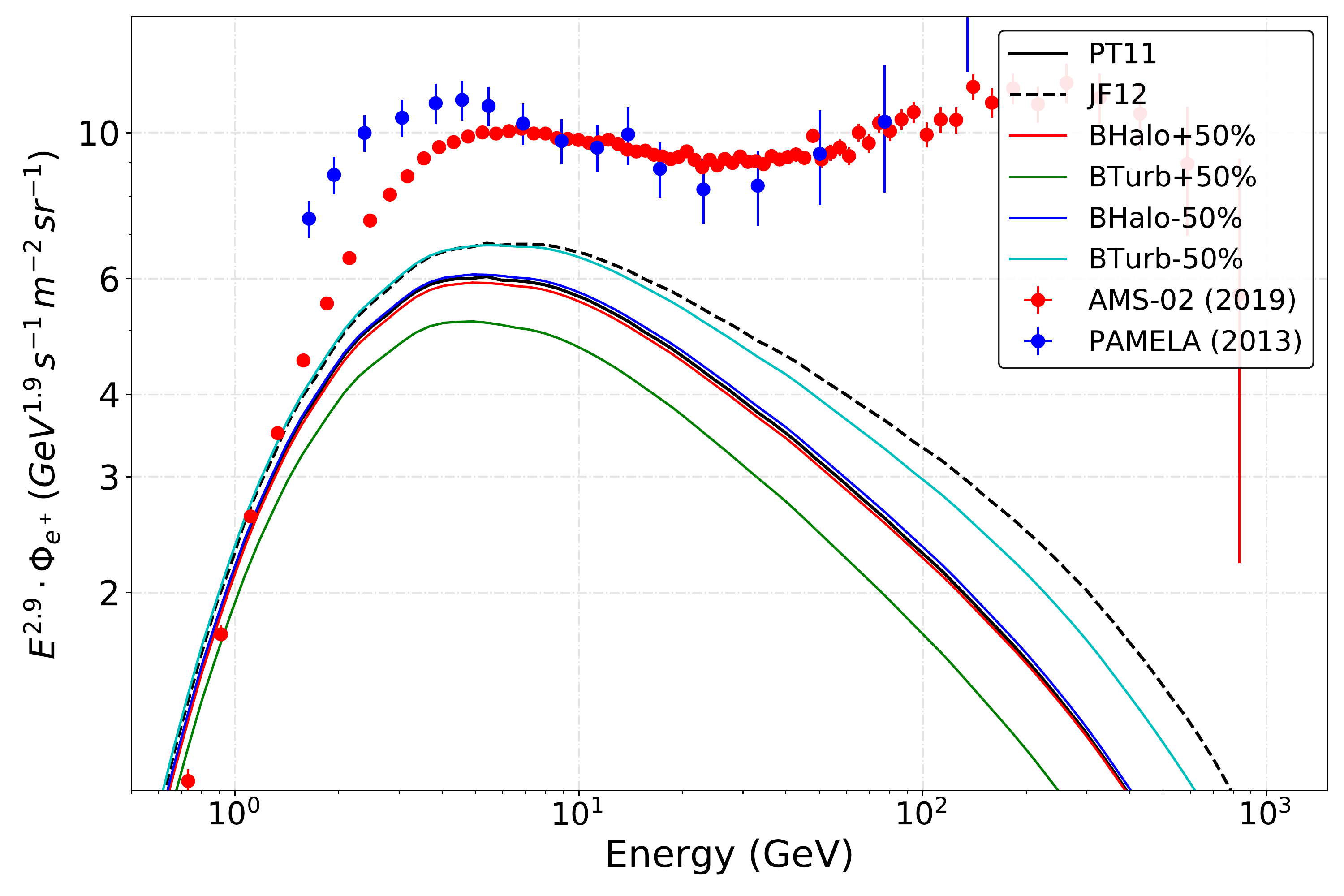}
\end{center}
\caption{Local positron spectrum evaluated from variations of $50\%$ of the intensity of the different components of the PT11~\cite{Pshirkov_2011} magnetic field model: $2\pm1$~$\mu$G for the disk component, $4\pm2$~$\mu$G for the halo component and $7.5\pm3.75$~$\mu$G. These are shown as solid lines. Moreover, the positron spectrum obtained using the JF12 model~\cite{Jansson_2012, Jansson_2012_2} adopted here is shown as a dashed black line. The left panel displays the predictions using the 2D configuration of the Galaxy, while the right panel shows those obtained using the 3D model.}
\label{fig:Blosses}
\end{figure} 

IC losses depend on the Galactic ISRFs, typically divided into the CMB, infrared, optical and (various) UV fields. Optical and UV fields that are produced from nearby stars are those affected by the largest uncertainties.
Ref.~\cite{Weinrich_halo} estimated that the uncertainties in the ISRF lead to an error in the estimation of the positron flux smaller than $5\%$ (at $E<100$~GeV), by comparing different models taken from Ref.~\cite{Delahaye_PositronXSs}. In turn, the uncertainties in the evaluation of the positron spectrum associated to synchrotron losses were estimated in Ref.~\cite{Weinrich_halo} to be the most important of these, reaching more than $10\%$ error below $100$~GeV. 

In Fig.~\ref{fig:Blosses}, we show the effect of adopting reasonable variations of our default magnetic field configuration (PT11; denoted as a solid black line in both panels) for the intensity of disk and halo components ($2\pm1$~$\mu$G for the disk component and $4\pm2$~$\mu$G for the halo component) in the estimation of the local positron spectrum. We also show the impact of varying the RMS intensity of the turbulent field by $50\%$ ($7.5\pm3.75$~$\mu$G). We remark that the turbulent magnetic field is much less known, not only because the RMS field intensity is difficult to be precisely measured, but also because of the uncertainties on its spatial structure.
While varying by $50\%$ the intensity of the disk or halo components affects the predicted spectrum in no more than a $8\%$ in the GeV region, we observe changes of up to $\sim20\%$ in our predictions from a $50\%$ variation of the intensity of the turbulent field. The different spectra obtained from these changes are shown as solid lines in Fig.~\ref{fig:Blosses}, where the left panel represents the predictions using our 2D model of the Galaxy and the right panel the predictions from our 3D model. We emphasise that adopting the 3D model also changes the bremsstrahlung energy losses and the distribution of the ISRF, which are not azimuthally symmetric within this model.

A more recent evaluation of the Galactic magnetic field, performed by Ref.~\cite{Jansson_2012} (JF12), also includes a striated random (or ordered random) field aligned along a particular axis over a large scale. 
The turbulent component in this model~\cite{Jansson_2012_2} was evaluated from the study of the WMAP7 $22$~GHz synchrotron emission from the CR electron density maps obtained by Ref.~\cite{Strong_CREdensity}. The strength of the local turbulent magnetic field was predicted to be $3-5\mu$G with an uncertainty of $\sim3\mu$G (notice that this is more than the $50\%$ variation that we show in Fig.~\ref{fig:Blosses}). It is important to note here that the electron spectra used to fit the synchrotron emission in that work were obtained using a different magnetic field model, which can certainly lead to large systematic uncertainties.
We also report in Figure.~\ref{fig:Blosses} the positron spectrum obtained using the JF12 Galactic magnetic field model (dashed black lines), using the best-fit parameters found in their study for the regular magnetic field and the turbulent field as in the PT11 model with $B_0^{turb}=4\mu$G. As we see in both panels of this figure, the change of the magnetic field models that we use yields a change in the positron flux predicted that can be of up to $20\%$ at $\sim5$~GeV. A formal study of the synchrotron emission from the models presented in this work is left for an upcoming work. However, we remark that the effect of energy losses on the local positron spectrum mainly depend on the value of the local magnetic field while the spatial distribution of these fields are not very relevant for these studies. 




\section{Discussion}
\label{sec:Discussion}

In this paper we have reported a new set of cross sections for CR electron and positron production derived from {\tt FLUKA} and reviewed the critical ingredients affecting our current predictions of the secondary positron flux at Earth in the GeV range. we remark that our evaluations can be very valuable not only for the GeV-TeV energy range, but also for the study of the $511$~KeV signal in the Galaxy, which constitutes a subject of debate right now~\cite{siegert2023positron}. In particular, including all resonances below $1$~GeV in our cross sections (specially neutrons) and considering the 3D gas distribution and magnetic field in the Galaxy considerably change the expected positron flux in different regions of the Galaxy at such low energies. Moreover, we highlight that our calculations include the production of positrons from triplet pair production (TPP)~\cite{Anguelov_1999, Gaggero:2013eik}, that become also relevant below a few hundred MeV.

We observe, from the calculations shown above, that the most important source of uncertainty in the evaluation of the local positron flux is the determination of the halo height, but even for a fixed halo height the rest of uncertainties can change our predictions by up to a factor of $2$ at a few GeV (and more below the GeV). We remark here that, while variations of the halo height, the gas density distribution and the energy losses can yield just smooth changes in the shape of the positron spectrum or in its normalization, different models for the Solar modulation, cross sections and reacceleration can also yield additional features in the energy region below a few GeV. Moreover, we remind the reader that the estimation of these uncertainties are subject to the traditional framework where positrons are not accelerated in SNR shocks (which can change significantly the high energy part of their spectrum~\cite{Mertsch:2014poa}) or other more exotic mechanisms of production of these particles (see, e.g. Ref.~\cite{DESARKAR20211}).

The high uncertainties in our estimations of the local positron flux have direct implications in the current searches for dark matter and also in the evaluation of the production of positrons (and electrons) by nearby sources, namely SNRs and PWNs.
For what concerns DM searches, there have been a few recent studies using positrons to constrain the annihilation rate of leptophilic dark matter~\cite{John_2021, Cholis_Krommydas, Cholis_Hoover}. Our results indicate that the searches for DM features in the positron spectrum will be highly affected by the astrophysical uncertainties, specially when the signals expected are not sharp, as it is the case of the expected signal from a heavy mass WIMP or the signal generated in the $\tau \tau$ channel. 
In fact, using synthetic data to test the ability of their analysis to spot this kind of features, the authors of Ref.~\cite{Cholis_Krommydas} pointed out that they were unable to identify masses larger than $\sim30$~GeV in the $\mu \mu$ channel and totally unable to correctly identify any signal generated in the $\tau \tau$ channel, even though they did not consider reacceleration in their approach, which may significantly affect their results.
On top of this, we note that even sharp DM signals may be masked by the emission of more conventional sources of positrons, as PWNs or even SNRs (see Ref.~\cite{Fornieri:2019ddi}). 

\begin{figure}[!t]
	\centering
	\includegraphics[width=0.47\textwidth]{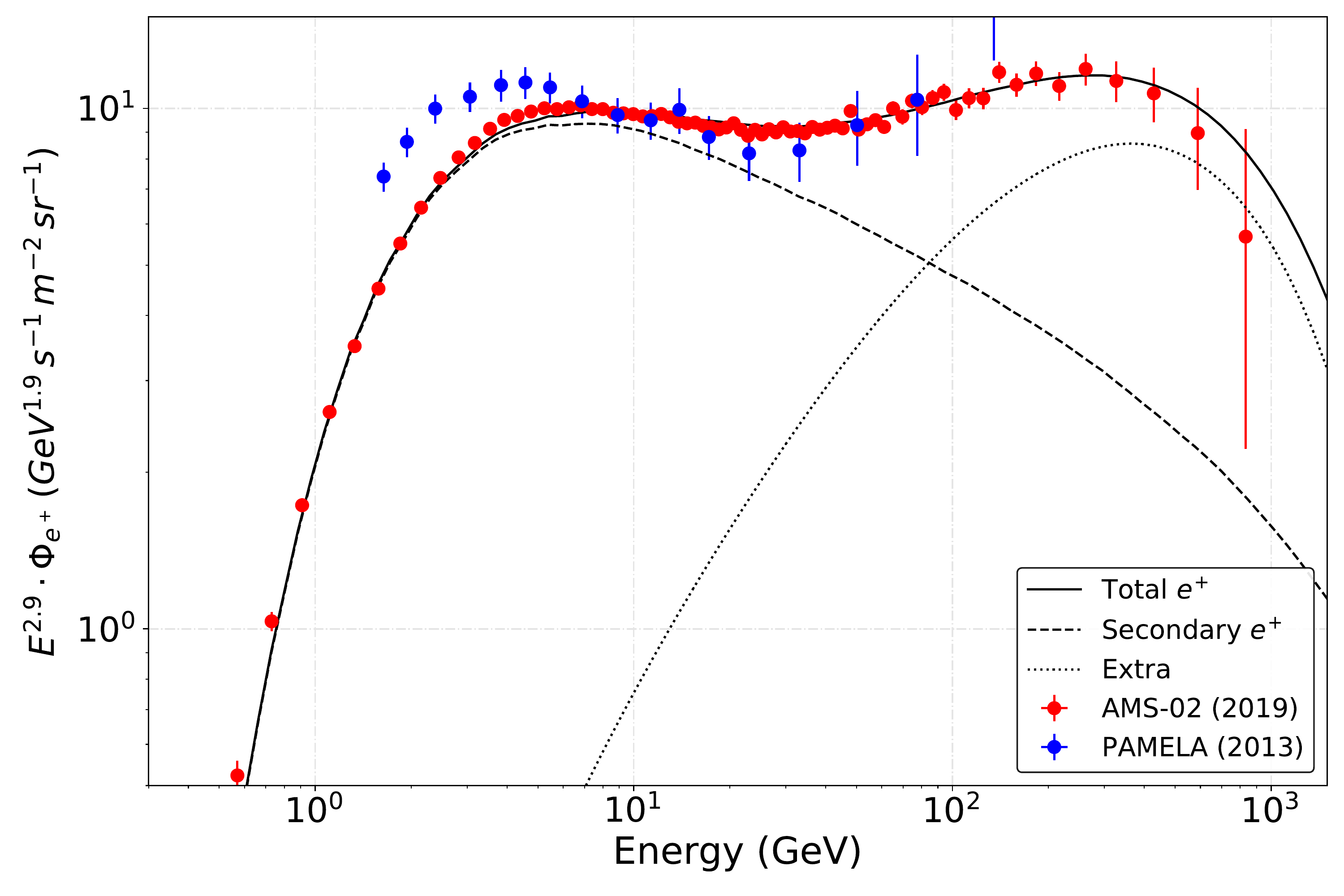} \hspace{0.2 cm}
	\includegraphics[width=0.47\textwidth]{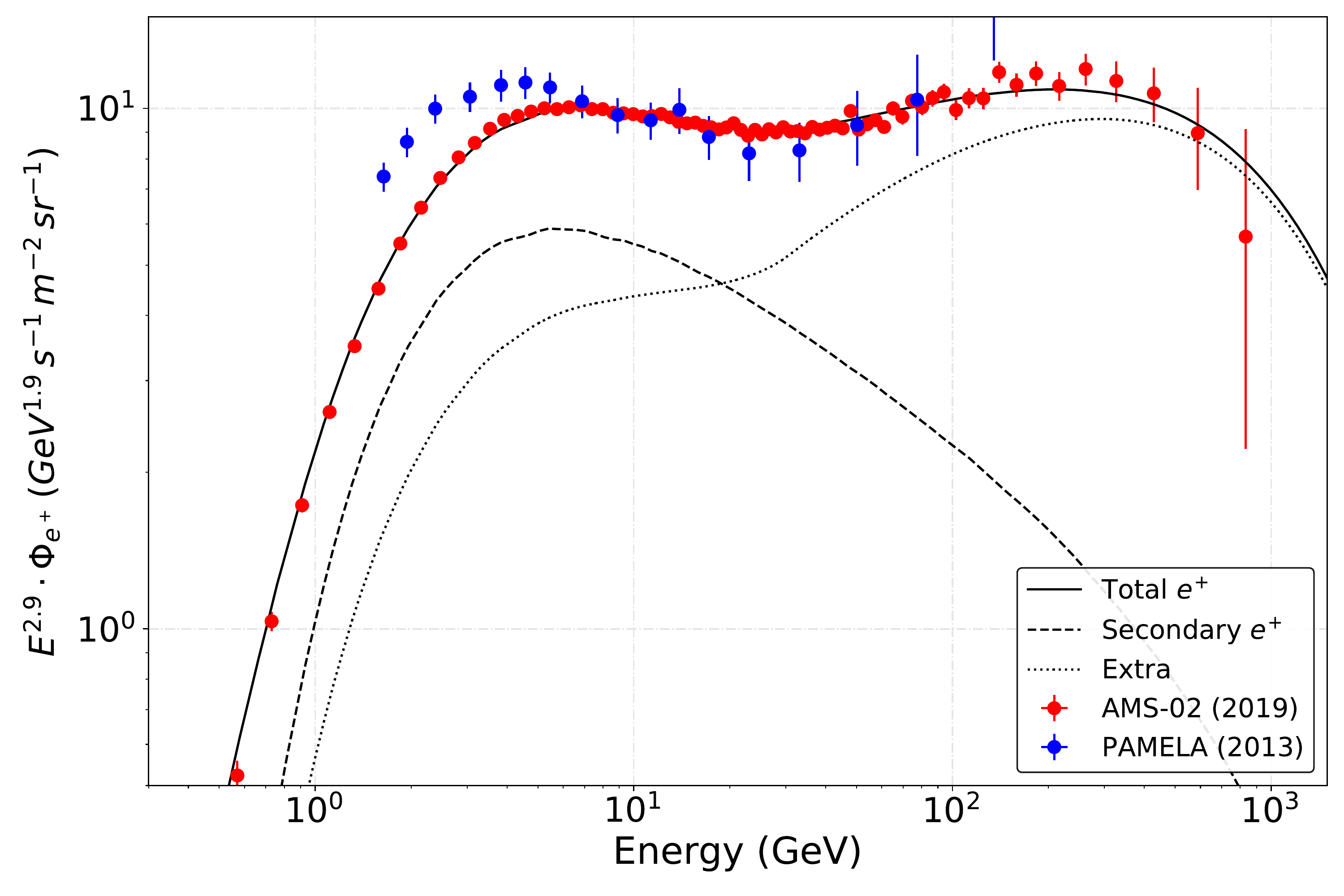}
	\caption{Local positron spectrum fitted to AMS-02 data for two different assumptions of the injection of positrons from PWN (dotted lines). In the left panel, we assume a single power-law for the injection of positrons and a secondary contribution (dashed line) adjusted to reproduce the low energy part of the spectrum. In the right panel, the secondary positron contribution is fixed to that found from the best-fit parameters and the positron injection is described by a broken power-law adjusted to the data.}
	\label{fig:Pos_Extrafit}
\end{figure}

Finally, we test the possibility of reproducing the AMS-02 data with a reasonable set of propagation parameters and astrophysical ingredients, adding an effective extra contribution representing the PWN injection of electrons and positrons. In a first model (left panel of Fig.~\ref{fig:Pos_Extrafit}), we parameterise this extra source by a distribution of sources following the SNR spatial distribution, that inject electrons and positrons evenly, following a single power-law in energy with a cut-off set at $1$~TeV. In this case, we find that we are able to reproduce the AMS-02 spectrum adopting a halo height of $4$~kpc, the FLUKA cross sections, the JF12 model adopted here, the Steiman-Cameron et al. 3D structure of the Galaxy (with $\sigma=0.5$~kpc) and the propagation parameters obtained from the B/C fit (with $D_0$ scaled to reproduce the B/C spectrum for a $4$~kpc halo height). The effect of Solar modulation is evaluated using the modified force-field approximation model described above, with $\phi=0.56$~GV and $\phi_n=0.15$~GV. We obtained a spectral index of the power-law of $1.45$, which is in good agreement with what is observed from IC emission around PWN.

In a second model (right panel of Fig.~\ref{fig:Pos_Extrafit}), we fix the halo size and propagation parameters to those found in the analysis of the B/C ratio (where the halo size is $H=7.5$~kpc) and use the FLUKA cross sections with the PT11 Galactic magnetic field model described above and the 3D distribution of the Galaxy ($\sigma=0.5$~kpc). We find that the Solar modulation parameters that allow us to fit the low energy part of the positron spectrum are $\phi=0.56$~GV and $\phi_n=0.25$~GV. In this case, we need to use a broken power-law for the injection of primary positrons in order to reproduce the AMS-02 data, where the spectral index is $2.5$ below $25$~GeV and $1.74$ above that energy. The difference in the spectral index found at high energies with respect to the single power-law case is due to the larger secondary positron contribution obtained at higher energies in the single power-law case.

What is remarkable from the evaluations shown in Fig.~\ref{fig:Pos_Extrafit} is that the secondary spectrum produced at high energies from these models is significantly higher than earlier estimations using the Kamae cross sections without taking into account the production of heavy species and the 2D model of the Galaxy. This is very relevant for the studies trying to adjust the average injection parameters from pulsars, since they depend on the assumed secondary positron flux in their fits (see Ref.~\cite{Orusa_2021}). In particular, in the case where the PWN contribution is represented by a single power-law, the secondary positron production is the dominant contribution up to $\sim70$~GeV, which is way higher than the recent expectations. However, we note that a single-power law for the description of the PWN contribution likely leads to an underestimation of the injection of pulsars at low energies (see Ref.~\cite{Manconi:2020ipm}). 
In this aspect, the case where the primary contribution of positrons is described by a broken power-law seems more credible, besides the fact that we are using the best-fit parameters of our model here. However the primary contribution obtained in this case is slightly higher than usual estimations. This indicates that these estimations likely represent reasonable upper and lower bounds for the secondary positron flux and the real secondary positron spectrum will lay somewhere in between.  
These spectra of electrons, positrons, H and He at Earth are available at the URL \url{https://github.com/tospines/Analyses-and-plotting-codes/tree/main/Fluka2021/PosEl_maps} in the form of fits files and the full maps with the distribution of these particles in the Galaxy are available upon request.


During the elaboration of this manuscript, a paper aiming at evaluating the local positron flux appeared in the ArXiv~\cite{dimauro2023novel}. This paper uses the Orusa et al.~\cite{Orusa} cross sections and estimates the positron spectrum using different models of diffusion and considering production of positrons up to Si. They adopt a 2D model of the gas distribution in the Galaxy and a simplified model for the Galactic magnetic field that allow them to study local energy losses. 
It is also interesting to note that most of their models lead to a spectral index of the diffusion coefficient $\delta \sim0.65$, which is significantly larger to the one found from previous analyses. In turn, they find that a model with a low-energy break in the injection spectrum of primary CRs lead to $\delta=0.39$, which is consistent with our findings, although they need too large cross sections scalings in this case.

\section{Summary and conclusions}
\label{sec:Conclusion}

The measurements of CR positrons enclose important information about the properties of the Milky Way and have the potential to reveal the existence of physics beyond the Standard Model. In this work we have investigated the main uncertainties affecting our predictions on the local positron spectrum. Moreover, we have employed for first time the new set of cross sections for production of electrons and positrons derived from the {\tt FLUKA} code~\cite{delaTorreLuque:2022vhm}, which uses optimised models of particle interactions updated with current accelerator data in order to improve the current calculations for CR interactions from the MeV to above the TeV.

We have reported novel cross sections for positrons and electrons computed using the {\tt FLUKA} code and compared them with the most widely used cross-section data-sets for different interaction channels.
Our evaluations indicate that cross sections uncertainties for p-p cross sections are definitely below $35\%$, for positrons, and slightly larger for electrons, comparing data sets that use very different approaches from a few hundred MeV to 10 TeV. Moreover, we find that the energy dependence of the predicted cross sections for p-p interactions is very similar below $100$~GeV. The relative difference in the predicted cross sections for the p-He channel are also below $\sim25\%$ for positrons and more than $50\%$ for electrons, although here we find significant differences in the energy dependence of the data-sets.
For He projectiles, the discrepancy found is never higher than $40\%$ for positrons, but higher than $100\%$ for electrons. Likewise, the discrepancies found in channels involving projectiles heavier than He are at the level of $60-100\%$ ($>100\%$) for positrons (electrons).

Then, we have investigated the effect of different diffusion parameters and the halo height in the predicted local positron spectra. We see that our estimations for two quite different propagation models yield very similar predictions above a few GeV, but at lower energies the effect of reacceleration severely changes the expected flux. On top of this, as the mean free path of positrons is shorter than other species, they are very affected by the height of the Galactic halo, which is not well constrained, and results in a factor $\sim2$ uncertainty in our determination of the positron flux at $5$~GeV and larger below. Interestingly, we show that the effect of reacceleration is mitigated for large values of H, as the diffusion coefficient gets larger to match observations on secondary-to-primary ratios (and vice versa).
In fact, it seems that the combination of low halo height and large $V_A$ is poorly compatible with the AMS-02 data. 
However, we also show that the effect of Solar modulation can also hide features related to reacceleration in the positron spectrum. Moreover, our estimations commonly use the simple force-field approximation, that can lead to large errors in our predictions at low energies. We find that the uncertainties in the treatment of Solar modulation may be the most important source of uncertainties below $2$~GeV.

Another important issue present in most of the recent evaluations of the positron spectrum is the use of an azimuthally symmetric gas density distribution (i.e. normally our models do not consider the spiral structure of the Galaxy). We discuss that this leads to errors in the estimations of the positron (and electron) spectrum above the GeV due to the scale length of the energy losses that these particles suffer. In our evaluations, we demonstrate that the use of a realistic gas density distribution may lead to up to a $30\%$ increase in the estimated flux at around $5$~GeV. Together with this, we report that variations of the intensity of the different components of the magnetic field can reasonably lead to $\sim10\%$ uncertainties in the $10$~GeV region, and that the main component affecting this is the turbulent field. Remarkably, we show that the uncertainties on the strength of the turbulent magnetic field can yield more than a $20\%$ variation in the predicted positron spectrum at $5$~GeV.

Finally, we explored models that allow us to reproduce the AMS-02 data including a simple primary source of positrons and electrons representing the injection from PWN. We tested two models for the secondary positron contribution: one where the secondary positron flux fits the local positron spectrum below a few GeV and the PWN injection spectrum is a simple power-law function, and the other where the secondary positron flux is the one found from our best-fit propagation parameters and halo height and where the PWN injection spectrum is described by a broken power-law. We discuss that these two models can represent upper and lower bounds of the expected secondary flux.
We remark that the secondary spectrum obtained at high energies from these models is significantly higher than that found using the Kamae cross sections without taking into account the production of heavy species and the spiral arm distribution of the interstellar gas in the Galaxy. 
On top of this, we have discussed what are the implications of the uncertainties in the evaluations of the positron spectrum for the current DM searches or the study of their injection from pulsars. 

We conclude that, in order to have more robust estimations of the local positron spectrum we need to have a more robust determination of the Galactic halo height and of the synchrotron energy losses, something that could be simultaneously achieved from studies of synchrotron radio emission. Then, we also emphasize the importance of considering the contribution from heavy CRs and accounting for the Galactic spiral arm distribution of sources and gas to produce more realistic evaluations of the local electron and positron spectrum.

\acknowledgments

P. De la Torre is supported by the Swedish National Space Agency under contract 117/19.
We acknowledge the {\tt FLUKA} collaboration for providing and supporting the code.
This work has been partly carried out using the RECAS computing infrastructure in Bari (\url{https://www.recas-bari.it/index.php/en/}). A particular acknowledgment goes to G. Donvito and A. Italiano for their valuable support. This project used computing resources from the Swedish National Infrastructure for Computing (SNIC) under project Nos. 2021/3-42 and 2021/6-326 partially funded by the Swedish Research Council through grant no. 2018-05973.


\appendix

\section{Diffusion parameters}
\label{sec:appendixA}

\begin{figure}[!b]
\centering
\includegraphics[width=0.6\textwidth,height=6.5cm]{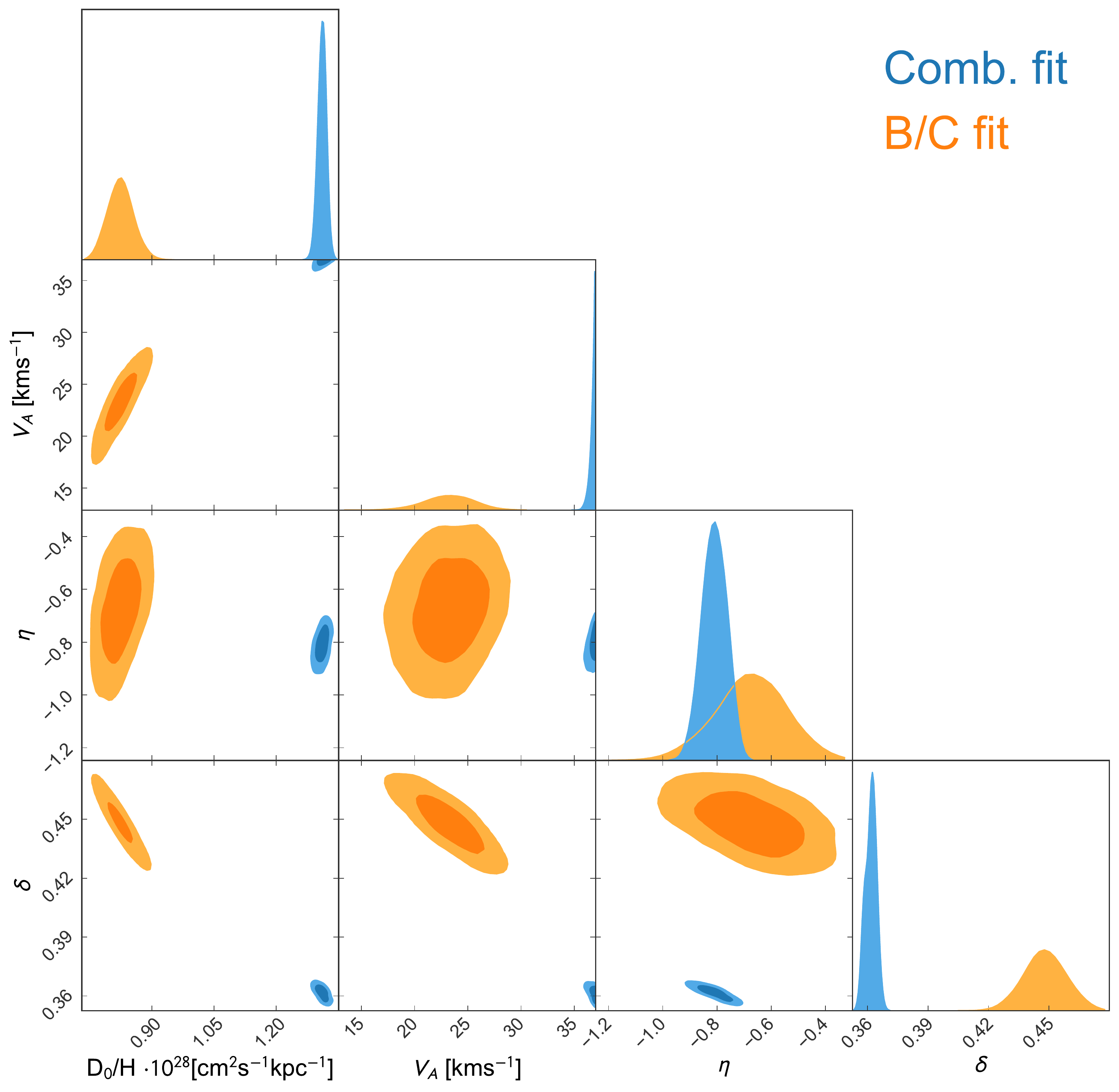} 
\caption{Corner plot comparing the PDFs of the two sets of propagation parameters used in this work. These parameters were obtained in Ref.~\cite{delaTorreLuque:2022vhm}, for a combined fit of the flux ratios of B, Be and Li to C and O (in blue) and the fit to B/C AMS-02 data (in orange). While the upper panels in every row represent the actual PDFs for each parameter, the 2D contours represent the $1$ and $2\sigma$ uncertainty in the determination of each pair of parameters.}
\label{fig:Corner_plot}
\end{figure}
In this appendix, we show a corner plot representing the two sets of propagation parameters used in this work. Figure~\ref{fig:Corner_plot} illustrates the probability distribution function (PDF) for each parameter found from the combined analysis of the flux ratios of B, Be and Li to C and O (in blue) and the fit to B/C AMS-02 data (in orange), as described in Ref.~\cite{delaTorreLuque:2022vhm}. This comparison clearly shows how different both sets of propagation parameters are, serving well to study how different propagation parameters affect our evaluations of the positron spectrum at Earth. In this figure, the contours represent the $1$ and $2\sigma$ uncertainty in the determination of each pair of parameters.
The best-fit parameters found from the fit to the B/C ratio are: $D_0=6.5$~cm$^2$/s, $\eta=-0.53$, $\delta=0.43$, $V_A=25.8$~km/s, while the parameters obtained in the combined fit performed in our companion work are: $D_0=9.8$~cm$^2$/s, $\eta=-0.81$, $\delta=0.36$, $V_A=39.7$~km/s. These propagation parameters allowed a simultaneous reproduction of not only all ratios of B, Be and Li, but also of the ratio $^3$He/$^4$He.

We are using the halo height value found from a fit the the $^{10}$Be ratios was $7.5$~kpc in our companion work, which is compatible to other recent estimations~\cite{Weinrich_halo, CarmeloBeB}.
As we see, these analyses yield quite different propagation parameters, allowing us to understand the impact of the transport uncertainties in our predictions for the local positron flux.

\section{Cross sections contributions}
\label{sec:AppendixB}

\begin{figure}[!h]
\centering
\includegraphics[width=0.47\textwidth]{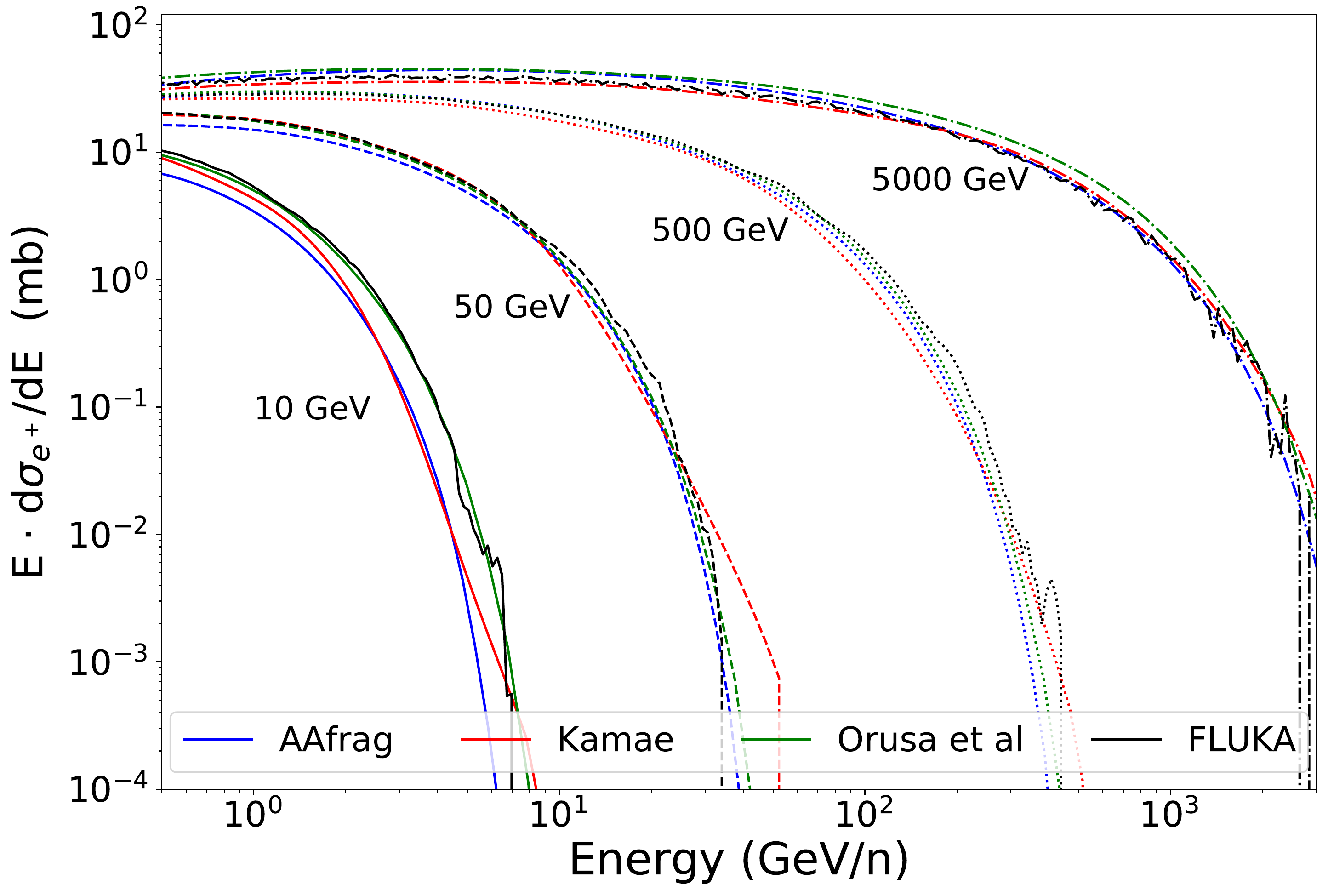} \hspace{.5cm}
\includegraphics[width=0.47\textwidth]{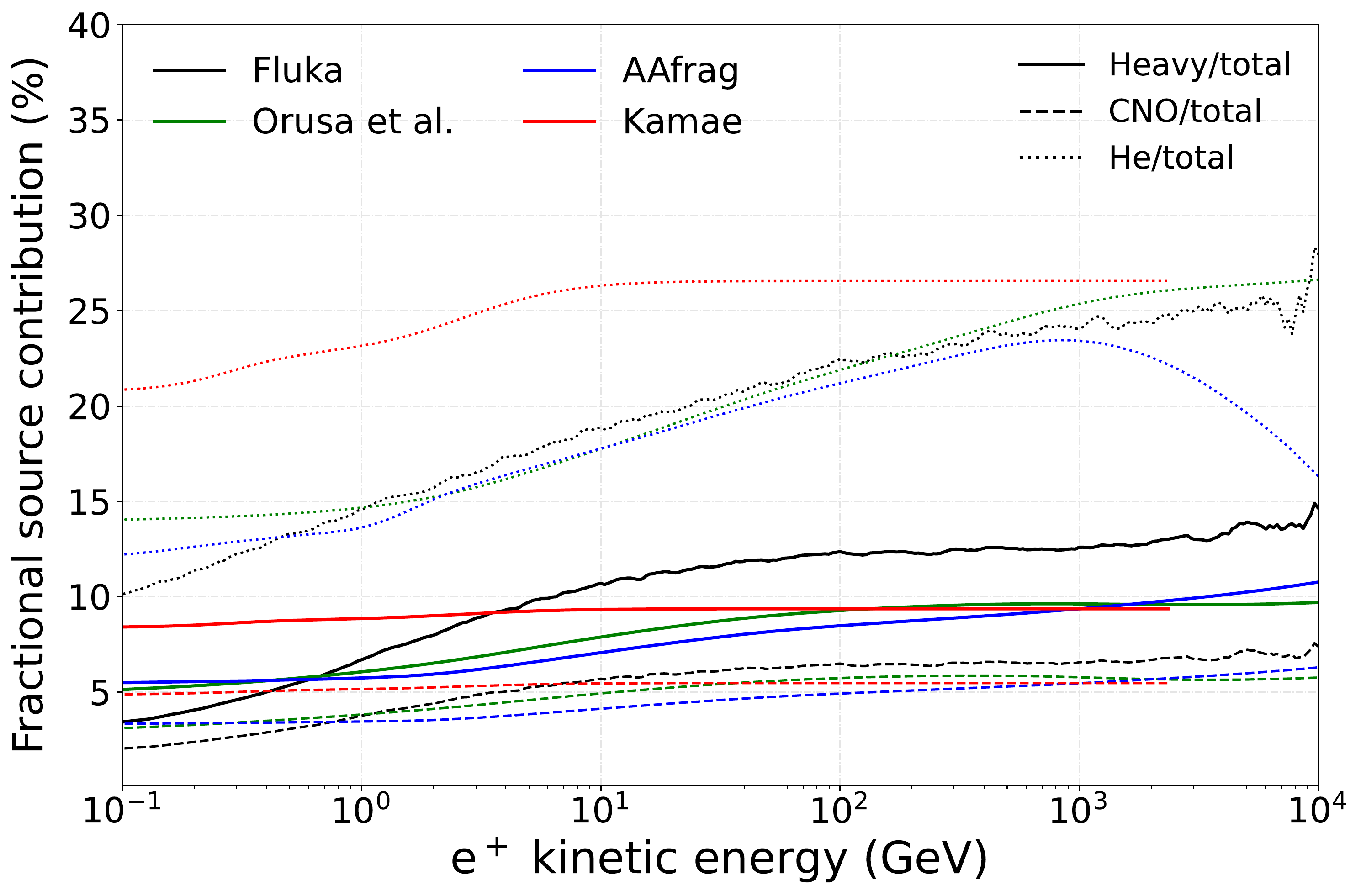}
\caption{\textbf{Left panel:} Differential inclusive positron cross sections for projectile energies of $10$ (solid lines), $50$ (dashed lines), $500$ (dotted lines) and $5000$~GeV (dash-dotted lines) for the FLUKA, Kamae, AAfrag and Orusa et al data-sets.
\textbf{Right panel:} Fraction of the total source term for the He contribution (dotted lines) the sum of the C, N and O contributions (dashed lines) and for the sum of all nuclei heavier than He (solid lines) for the different data-sets discussed in this work compared to the FLUKA predictions. For Kamae, AAfrag and Orusa et al, in those interaction channels where the cross sections are not available, the p-p cross sections have been scaled following the $A^{0.9}$ relation.}
\label{fig:He_contrib}
\end{figure}

In this appendix, we provide some additional details above the cross section for production of CR electrons and positrons derived from FLUKA. In the left panel of Fig.~\ref{fig:He_contrib}, we show the differential inclusive positron cross sections at different projectile energies ($10$, $50$, $500$ and $5000$~GeV) for the FLUKA, Kamae, AAfrag and Orusa et al data-sets. We find these estimations to be in agreement within $\sim20-40\%$ for the p-p channel.
In the right panel of Figure~\ref{fig:He_contrib}, we illustrate the fractional contribution to the total positron source production of He (dotted lines), the sum of the C, N and O source term contributions (dashed lines) and the sum of the source terms of all nuclei heavier than He (solid lines). Here, we observe a remarkable agreement between the estimations for the He fractional contribution from the Orusa et al, AAfrag and FLUKA cross sections.

Moreover, we have also tested the scaling of the AAfrag and FLUKA positron cross sections with the mass number $A$, fitting the ratio of the cross sections of interaction of a nucleus of mass number $A$ with a proton over the cross sections for the p+p interaction as $A^{s}$. We find that both the AAfrag cross sections and the FLUKA cross sections lead to $s\sim0.9$, which is quite coincident to the one found from the analysis of experimental data~\cite{Orusa}.

\begin{figure*}[p]
    \centering
    \includegraphics[width=0.49\textwidth]{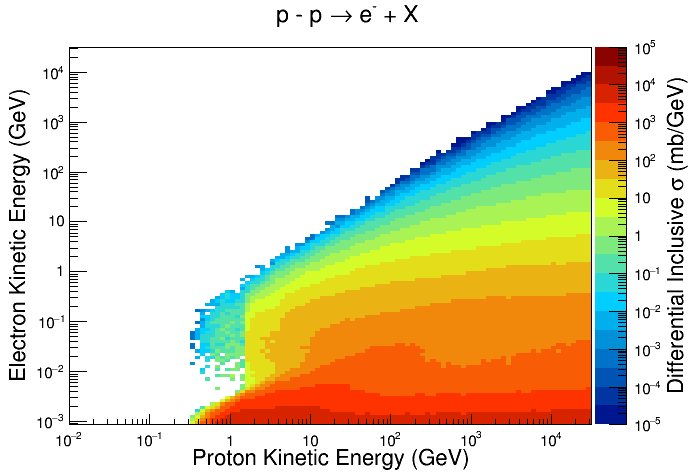}
    \includegraphics[width=0.49\textwidth]{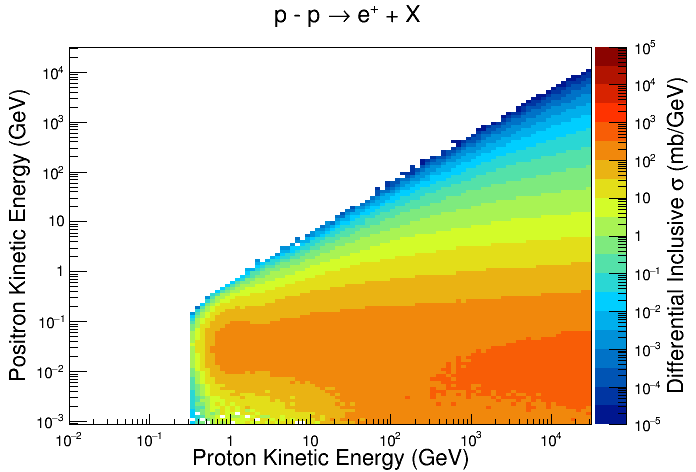}
    \includegraphics[width=0.49\textwidth]{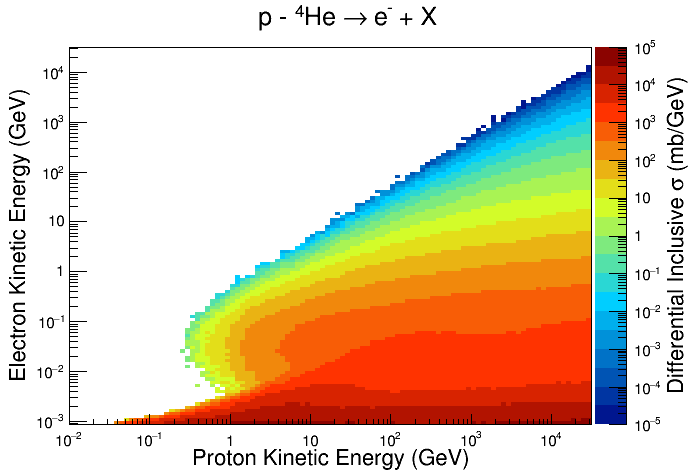}
    \includegraphics[width=0.49\textwidth]{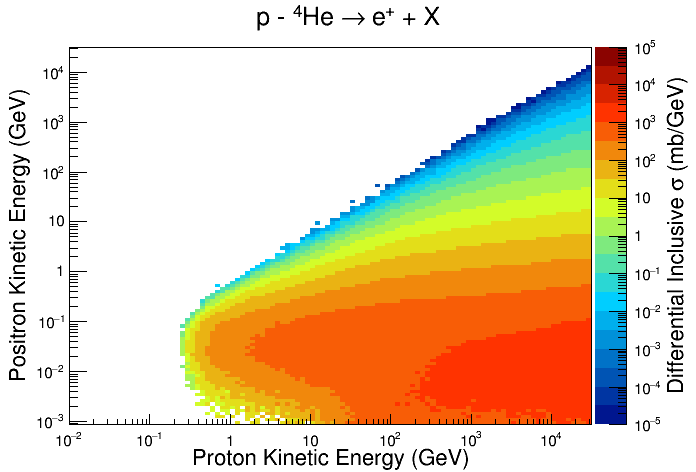}
    \includegraphics[width=0.49\textwidth]{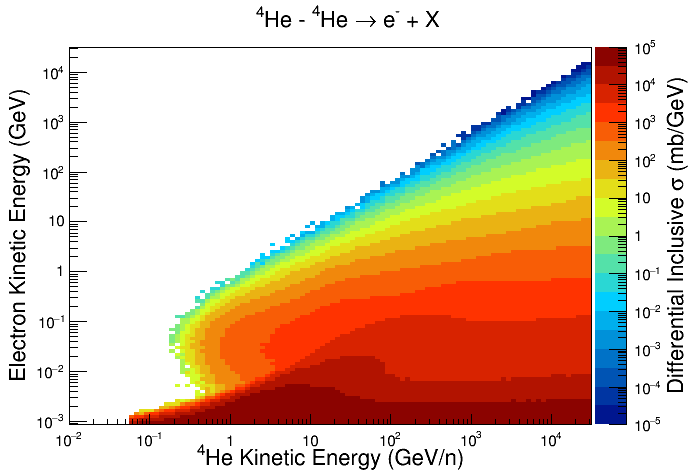}
    \includegraphics[width=0.49\textwidth]{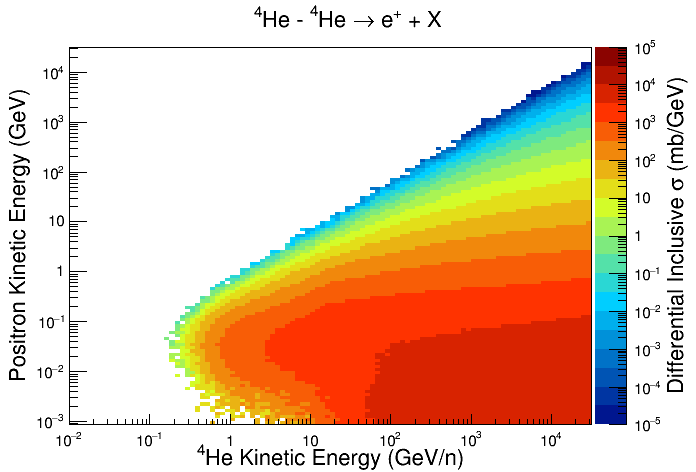}
    \includegraphics[width=0.49\textwidth]{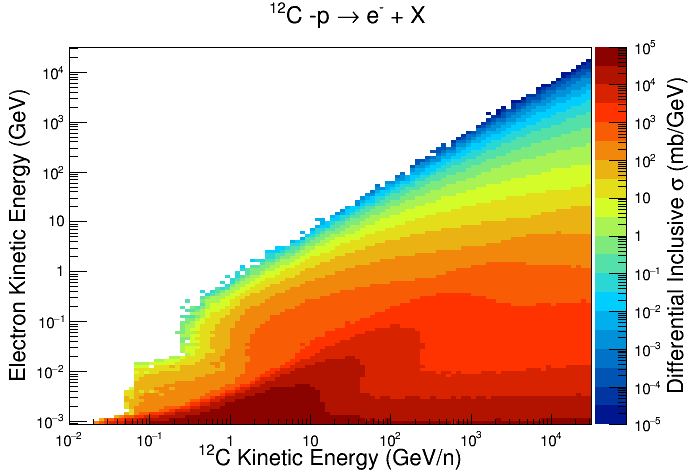}
    \includegraphics[width=0.49\textwidth]{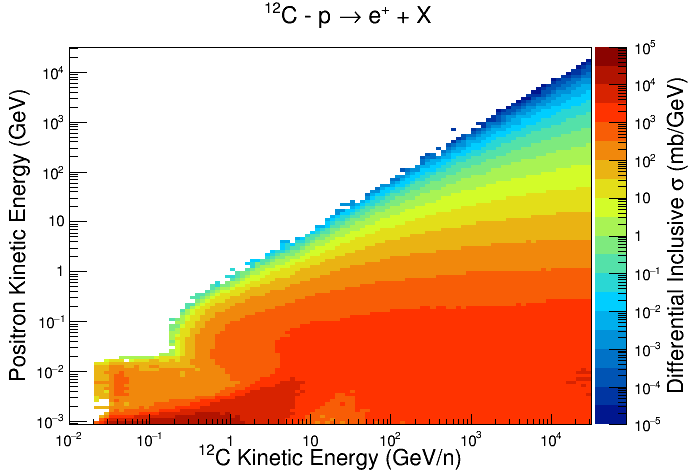}
    \caption{Differential cross sections of electron (left column) and positron (right column) production as function of the energy of the projectileand the daughter particle, obtained from {\tt FLUKA}, for p-p, p-$^4$He, $^4$He-$^4$He and $^{12}$C-p collisions.}
    \label{fig:2Delepos}
\end{figure*}

\begin{figure*}[ph]
    \centering
    \includegraphics[width=0.49\textwidth]{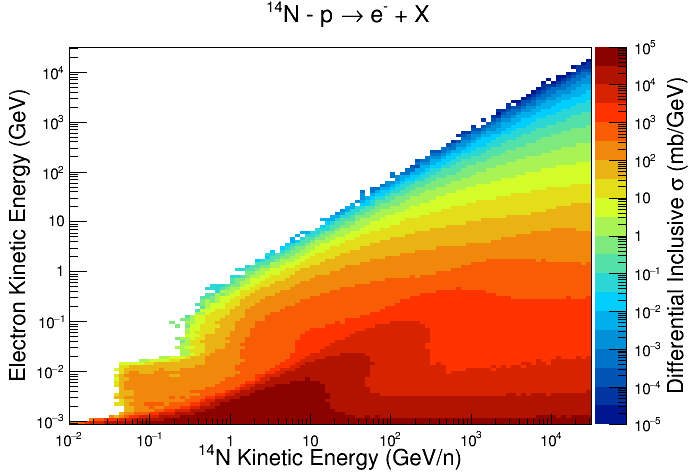}
    \includegraphics[width=0.49\textwidth]{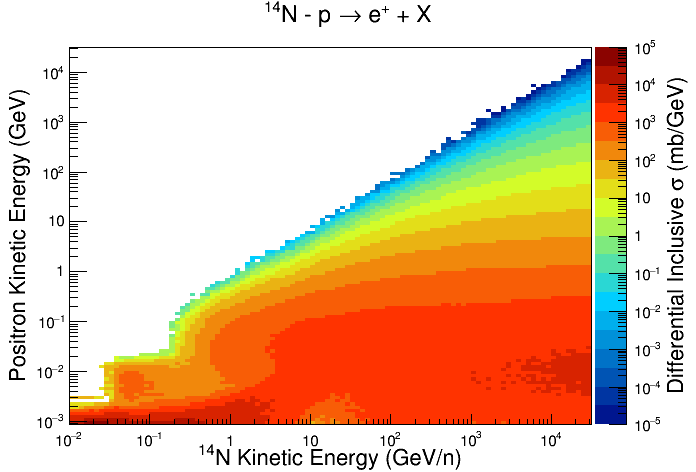}
    \includegraphics[width=0.49\textwidth]{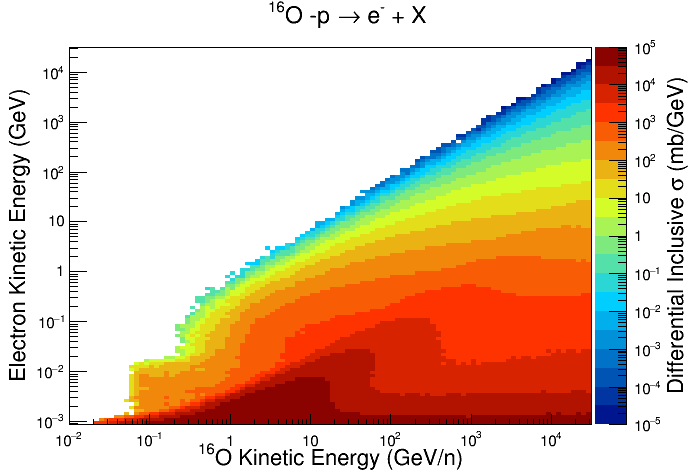}
    \includegraphics[width=0.49\textwidth]{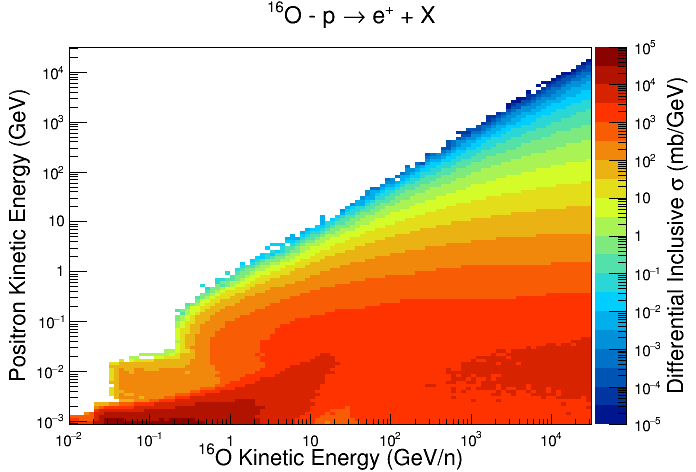}
    \includegraphics[width=0.49\textwidth]{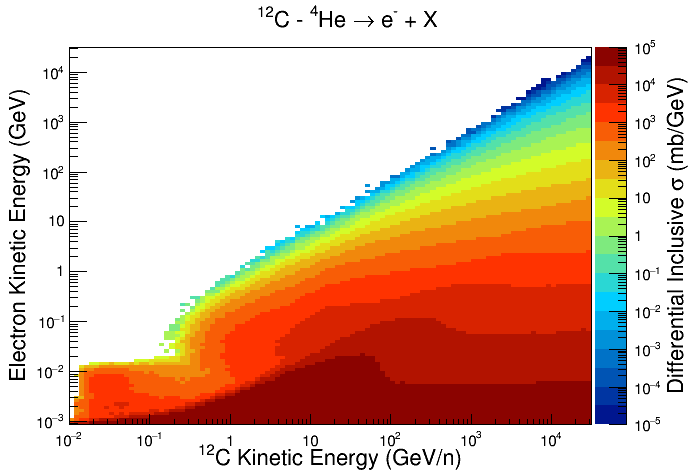}
    \includegraphics[width=0.49\textwidth]{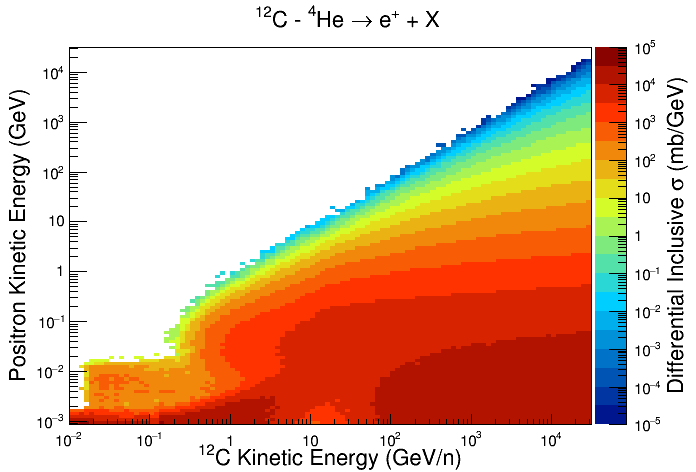}
    \includegraphics[width=0.49\textwidth]{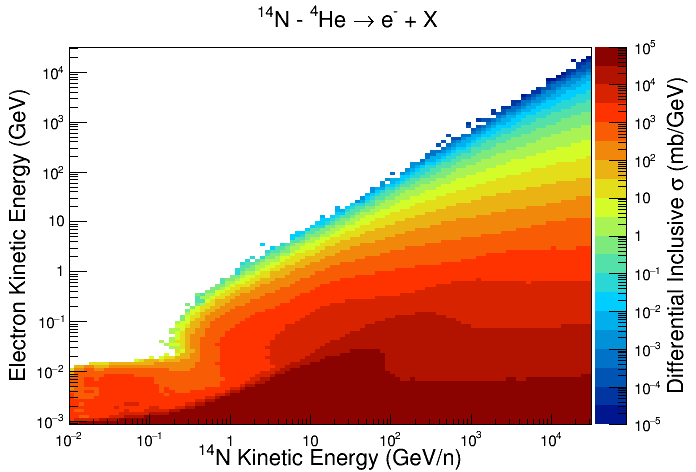}
    \includegraphics[width=0.49\textwidth]{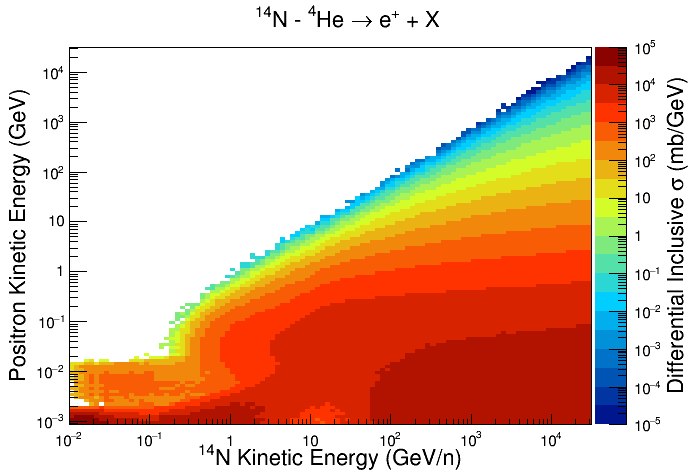}
    \caption{Similar to what is shown in Fig.~\ref{fig:2Delepos}, but for $^{14}$N-p, $^{16}$O-p, $^{12}$C-$^4$He and $^{14}$N-$^4$He collisions.}
    \label{fig:2Delepos_1}
\end{figure*}

Finally, we show the cross sections of production of electrons (Fig.~\ref{fig:2Delepos}) and positrons (Fig.~\ref{fig:2Delepos_1}) for each energy of the projectile CR and daughter particle. Here, we include every resonance incorporated in the {\tt FLUKA} code. We note here that, 
the decay of neutrons (antineutrons) constitute a significant fraction of the total electron (positron) yield below $\sim10$~MeV. 

\clearpage
\section{Lepton anisotropy}
\label{sec:Anisotropy}

\begin{figure}[!h]
\centering
\includegraphics[width=0.60\textwidth]{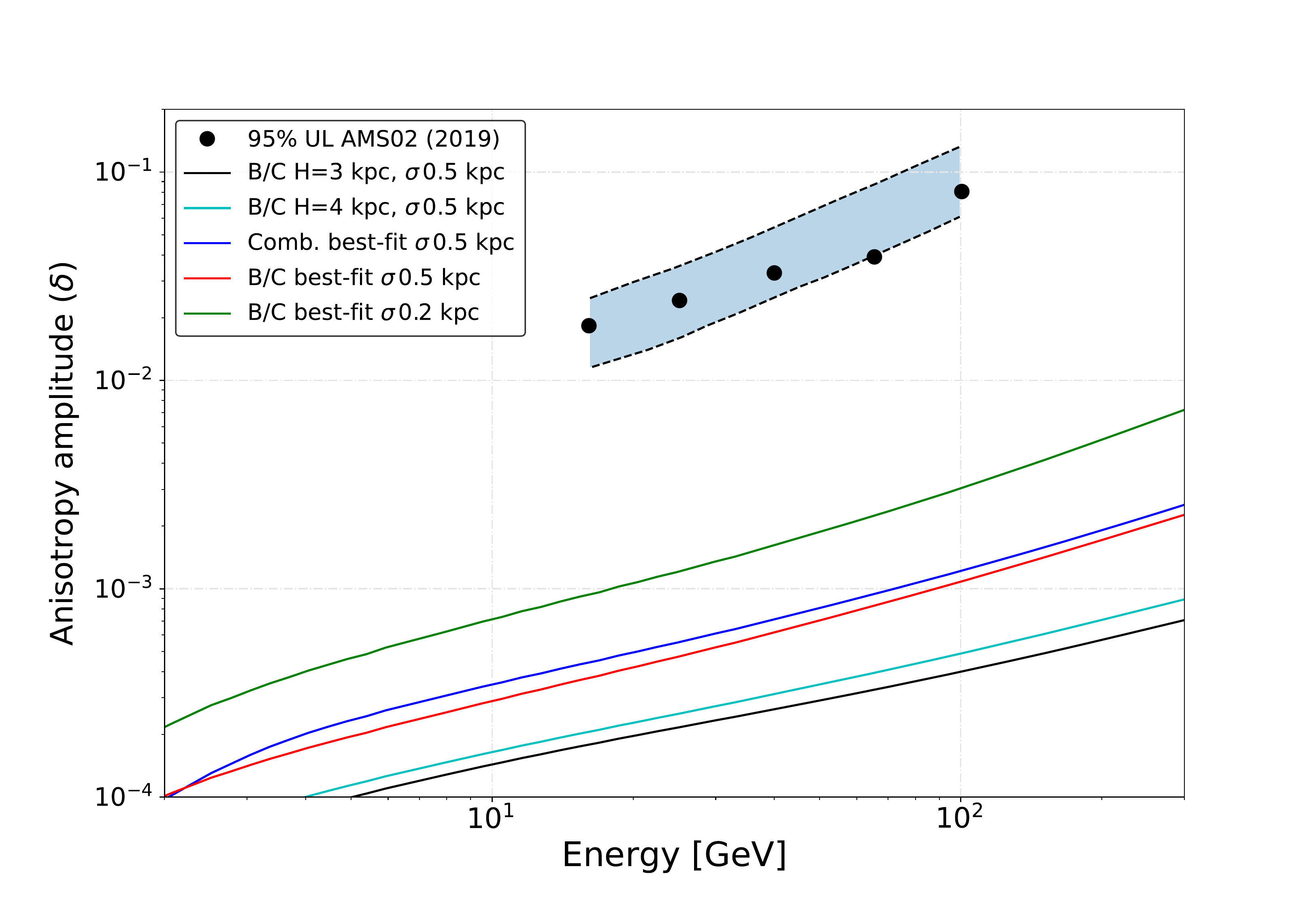}
\caption{Energy spectrum of the positron anisotropy amplitude predicted for different configurations of the the Galaxy structure explored in this work. We compare the predictions obtained using the 3D model of the Galaxy described above for the propagation parameters found from the B/C fit at different values of the arm width ($0.2$ and $0.5$~kpc) and for different halo height ($3$, $4$, and $7.5$~kpc). This is compared to the most recent the upper limits reported by the AMS-02 collaboration~\cite{Velasco_2020}.}
\label{fig:Anisotropy}
\end{figure}

Anisotropies in the arrival direction of electrons, positrons or protons have been searched from long time given that they would allow us to reveal signatures of nearby sources of positrons and even other unexpected phenomena. 
We report in Fig.~\ref{fig:Anisotropy} the predicted energy spectrum of the dipole anisotropy amplitude for different configurations of our setup, computed following the diffusive-regime approximation~\cite{Ginz&Syr}: $\delta = \frac{3D(E)}{c} \frac{|\nabla f(E, \textbf{r})|}{f(E, \textbf{r})}$, where $f(E, \textbf{r})$ is the positron density  and $D(E)$ is the spatial diffusion coefficient.  

In this figure, we show the predicted anisotropy amplitude for our 3D setup (which employs, by default, a halo height $H=7.5$~kpc) with arm width of $0.5$~kpc for the propagation parameters obtained from the B/C fit (red line) and obtained from the combined fit (blue line). In addition, we include the predicted amplitude spectrum for the same setup but with arm width of $0.2$~kpc (green line). Finally, we also report our predictions for two other different values of the halo height ($3$ and $4$~kpc).
Here, we find that both, the halo height and the arm-width have a very significant impact in our predictions: 
As we see, different arm width may imply a difference of more than a factor of $2$ in the predicted anisotropy amplitude, and the halo height seems to have a similar importance. However, these predictions still lay at least one order of magnitude below the upper limits recently reported by the AMS-02 collaboration~\cite{Velasco_2020}. 
In addition, we remind the reader that there are a few other ingredients that can change this prediction: the assumption that sources are smoothly distributed in the Galaxy, non-spatial dependence of the diffusion coefficient or reacceleration of positrons at SNR shocks could appreciably change this prediction.

\newpage
\section{Electron spectrum}

\begin{figure}[!h]
	\centering
    \includegraphics[width=0.52\textwidth]{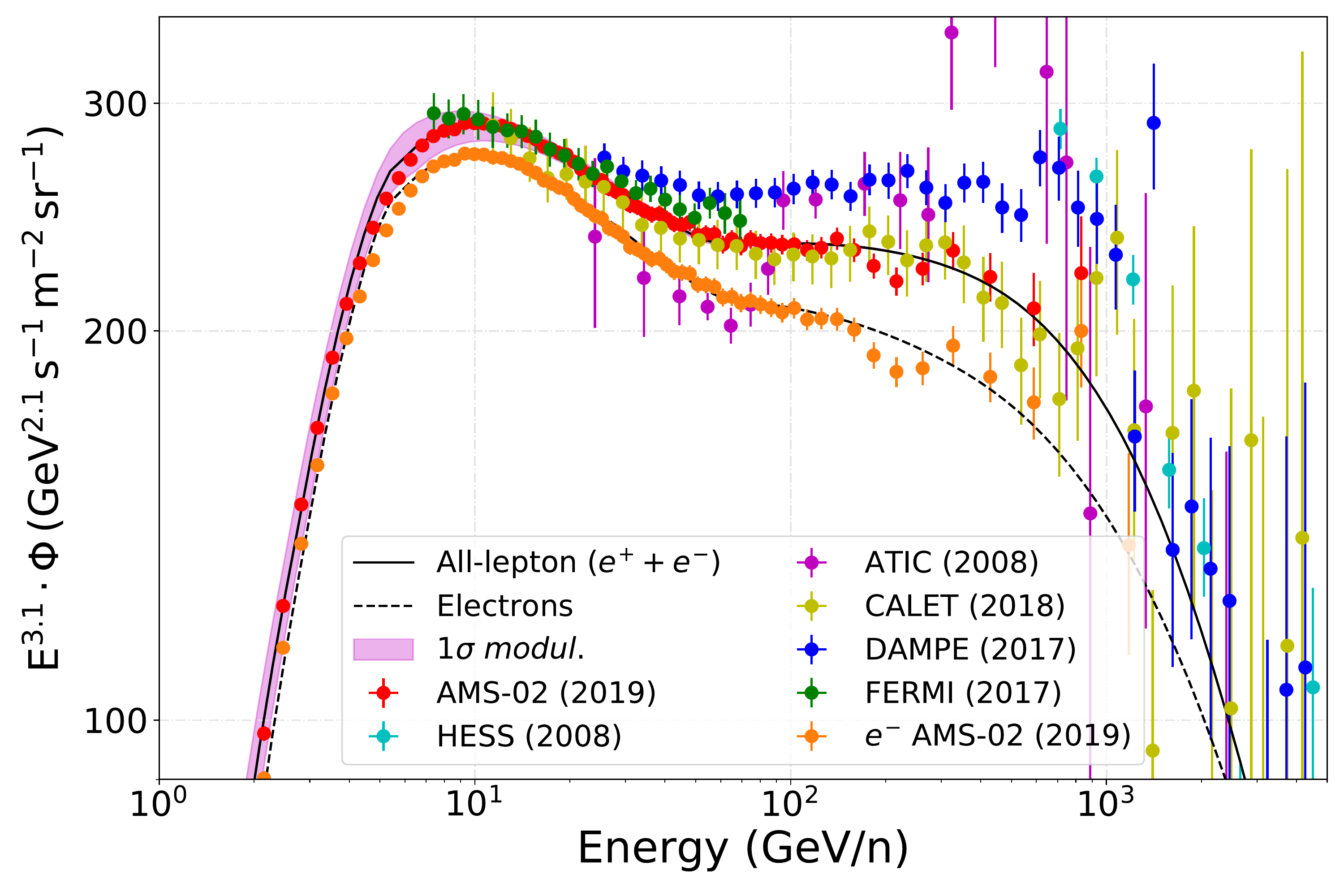}
	\caption{Predicted total CRE spectra compared to AMS-02, Fermi-LAT, DAMPE and CALET data in the case where the pulsar injection of electrons and positrons is adjusted as a broken power-law. The dashed line represents the predicted flux of electrons, fitting to the AMS-02 electron flux. The uncertainty band related to a variation of the Fisk potential of $\pm0.1$~GV is shown for the spectra in both figures.}
	\label{fig:El_spectrum}
 
\end{figure}

In Figure~\ref{fig:El_spectrum} we show the electron and electron+positron spectra fitted to reproduce AMS-02 data~\cite{Aguilar:2014mma} with the propagation parameters found from the B/C analysis, the PT11 Galactic magnetic field model described above, the 3D distribution of the Galaxy ($\sigma=0.5$~kpc) and using a Fisk potential $\phi = 0.58 \units{GV}$. The pulsar contribution of electrons and positron is the one obtained from the fit of a broken power-law, as explained in section~\ref{sec:Discussion} (right panel of Figure~\ref{fig:Pos_Extrafit}).
The electron and total CRE spectra are compared to data, from Fermi-LAT~\cite{Ackermann:2010ij}, DAMPE~\cite{Ambrosi:2017wek}, CALET~\cite{CALET_electrons} and AMS-02~\cite{Aguilar:2014mma}. 
The uncertainty band related to a variation of the Fisk potential of $\pm0.1$~GV is included for these spectra. 

\bibliographystyle{apsrev4-1}
\bibliography{biblio}

\end{document}